\documentclass{iucrjournals}

\usepackage{graphicx}%
\usepackage{multirow}%
\usepackage{multicol}
\usepackage{amsmath,amssymb,amsfonts}%
\usepackage{amsthm}%
\usepackage{mathrsfs}%
\usepackage[title]{appendix}%
\usepackage{xcolor}%
\usepackage{textcomp}%
\usepackage{manyfoot}%
\usepackage{booktabs}%
\usepackage{algorithm}%
\usepackage{algorithmicx}%
\usepackage{algpseudocode}%
\usepackage{listings}%
\usepackage{siunitx}
\usepackage{float}
\usepackage{lmodern}
\usepackage{microtype}

\title{Accurate Evaluation of Nanoscale Spatiotemporal Dynamics with Electron Correlation Microscopy}

\author[a]{Po-Cheng Kung}

\author[a]{Ajay Annamareddy}

\author[b]{Mark D.Ediger}

\author[a]{Dane Morgan}

\author[a]{Paul M. Voyles\IUCrCemaillink{paul.voyles@wisc.edu}}

\affil[a]{Department of Materials Science and Engineering, University of Wisconsin-Madison, Madison, Wisconsin, 53706, USA}

\affil[b]{Department of Chemistry, University of Wisconsin-Madison, Madison, Wisconsin, 53706, USA}

\begin{document}
\maketitle

\begin{synopsis}
A new physics-inspired analysis framework for nanobeam electron correlation microscopy eliminates systematic artifacts introduced by uncritically adopting photon correlation spectroscopy analysis methods. The proposed approach enables the accurate measurement of nanoscale spatiotemporal dynamics and correctly isolates stable crystalline phases from relaxing fluid domains in supercooled liquids.
\end{synopsis}

\begin{abstract}

Electron correlation microscopy (ECM) can measure materials dynamics with nanoscale spatial resolution from intensity correlation functions. However, adopting X-ray photon correlation spectroscopy (XPCS) normalization frameworks unchanged when calculating intensity correlations can introduce errors. Due to the constrained sampling volumes and larger speckle sizes in nanobeam electron diffraction, XPCS-style time-averaging and scattering-vector averaging introduce systematic artifacts, such as artificial anti-correlations or elevated baselines that lead to systematic errors in structural relaxation times and stretching exponents. This work presents physics-inspired, ECM-specific intensity normalizations over time- and azimuthal-averaged intensities of the first diffraction ring that limit those errors. The framework is validated using molecular dynamics simulations of a CuZr supercooled liquid to benchmark against the self-intermediate scattering function, successfully reproducing relaxation times. When applied to experimental time-resolved 4D STEM datasets of a Pt57.5Cu14.7Ni5.3P22.5 nanowire, the method correctly identifies highly stable, unchanging nanoscale crystalline phases that were erroneously misclassified as relaxing domains by previous frameworks. Other previous ECM research is reevaluated in light of these observation. This robust approach establishes an artifact-free pathway for evaluating localized spatiotemporal relaxation behaviors.
\end{abstract}




\section{Introduction}\label{intro}

As the temperature approaches the glass transition temperature ($T_g$), many supercooled liquids (SCLs) undergo qualitative changes in their dynamic behavior. With a drastic increase in the structural relaxation time, the dynamics become highly heterogeneous, with local dynamics that can vary by orders of magnitude across a few nanometers~\cite{Ediger2000,Tanaka2025,Berthier2011}. This behavior has been studied by dynamic hole burning experiments~\cite{Cicerone1995,Schmidt-Rohr1991,Schiener1997}, and inferred from the ensemble-averaged relaxation behaviors measured via mechanical spectroscopy~\cite{Hachenberg2006,Richert2010}, dielectric spectroscopy~\cite{Schonhals2003}, solid-state nuclear magnetic resonance spectroscopy~\cite{Reinsberg2001,Nunes2014}, and X-ray photon correlation spectroscopy (XPCS)~\cite{Ruta2012,Madsen2020,Das2019,Bikondoa2017,Jain2020,Conrad2015}. However, because these methods involve volume averaging, they lack the spatial resolution necessary for the direct observation of spatially heterogeneous dynamics in real space.

To achieve higher spatial resolution, researchers have utilized electron correlation microscopy (ECM)~\cite{Zhang2018,Zhang2017,Chatterjee2021,Nakazawa2025,Nakazawa2023Structure,Nakazawa2023,Huang2024,Vaerst2023,Spangenberg2021}. ECM uses intensity autocorrelation functions analogous to XPCS, but it achieves nanometer to sub-nanometer resolution through the use of electron optics. Early ECM was performed by tracking the intensity changes of a single scattering vector ($\mathbf{k}$) along the first diffraction ring via a time series of tilted dark-field (DF) transmission electron microscopy (TEM) images~\cite{Zhang2017,Zhang2018,Vaerst2023,Spangenberg2021,Chatterjee2021}. By estimating the structural relaxation time from the intensity autocorrelation function of each TEM image pixel, this technique observes spatially heterogeneous dynamics. However, because this method probes only a single scattering vector, it fails to capture the relaxation events of groups of atoms that scatter in different directions.

Recently, very high speed electron detection has made ECM based on time-resolved four-dimensional transmission electron microscopy (4D STEM) possible~\cite{Nakazawa2023,Nakazawa2023Structure,Huang2024,Nakazawa2025}. In time-resolved 4D STEM, two-dimensional nanobeam electron diffraction (NBED) patterns are collected at each spatial position repeatedly as a function of time. Therefore, the intensity autocorrelation functions can be calculated using the full range of scattering vectors to capture more dynamic events. Furthermore, by sampling over multiple scattering vectors, NBED-ECM allows for the measurement of the $\mathbf{k}$ ensemble-averaged momentary dynamics at each position. The momentary relaxation time describes the additional elapsed time required for a local structure to relax starting from any point in the time series of the measurement~\cite{Das2019}. NBED-ECM also enables the simultaneous observation of spatially and temporally heterogeneous dynamics.

ECM is largely based on XPCS, adopting many of its analysis tools and methodologies~\cite{Zhang2018,Zhang2017,Nakazawa2023,Nakazawa2023Structure,Nakazawa2025,Vaerst2023,Huang2024,Chatterjee2021,He2015}. However, despite both using a (partially) coherent source, photon and electron diffraction exhibit important differences, especially in the case of NBED. XPCS samples a large number of atoms using a broad beam, resulting in speckles that cover a very small range of diffraction space. NBED-ECM from one probe position samples the dynamics from a much smaller volume in real space than XPCS, and the speckles in reciprocal space are correspondingly larger. Thus, XPCS methods cannot be directly adopted, and methodologies for calculating the correlation functions specifically for NBED-ECM data are needed.  

Here, we propose a physics-based method for calculating correlation functions tailored to the NBED data. The proposed method is validated by comparing the structural dynamics estimated from simulated NBED-ECM data of a molecular dynamics (MD) trajectory of a CuZr metallic SCL to the trajectory's self-intermediate scattering function (ISF). The improved NBED-ECM method is also applied to experimental time-resolved 4D STEM datasets of a Pt\textsubscript{57.5}Cu\textsubscript{14.7}Ni\textsubscript{5.3}P\textsubscript{22.5} SCL nanowire, demonstrating the method's capability to study spatially and temporally heterogeneous dynamics in complex metallic SCL systems containing nanoscale crystal phases. Previous ECM results are reconsidered in light of the identified artifacts and revised correlation functions.

\section{Materials and Methods}

\subsection{Molecular Dynamics Simulation of CuZr Supercooled Liquids}

The model glass investigated in this study is an equiatomic Cu\textsubscript{50}Zr\textsubscript{50} metallic glass, a well-established binary glass-forming alloy. Atomic interactions were described using the embedded atom method (EAM) potential developed by Mendelev et al.~\cite{Mendelev2009}. The system was initially equilibrated in the liquid state at 2000 K for 2 ns and subsequently cooled, all under NPT conditions. Cooling from 2000 K to 1000 K was performed in 50 K increments at an effective rate of 100 K per 6 ns, followed by further cooling in 20 K increments at an effective rate of 100 K per 60 ns. The glass transition temperature ($T_g$) was determined to be approximately 700 K from the temperature dependence of the system volume. 

The present work focuses on the dynamics of the supercooled liquid (SCL) state at 760 K. The configuration obtained at this temperature during the cooling protocol was used as the initial state for subsequent production simulations performed in the NVT ensemble. No measurable drift in the potential energy was observed during these runs, confirming that the system remained in equilibrium within the SCL regime. All molecular dynamics simulations were carried out using the LAMMPS package~\cite{Plimpton1995}. Structural and dynamical analyses were performed using inherent structures obtained by minimizing instantaneous configurations with the conjugate-gradient algorithm implemented in LAMMPS. The simulation cell contained 16,384 atoms, periodic boundary conditions were applied in all three spatial directions, and a time step of 1 fs was employed throughout.

The MD trajectory of the CuZr SCL was sampled every 20 ps to obtain 501 snapshots. 4D STEM data was simulated for each snapshot using a multislice algorithm through the abTEM Python package~\cite{madsen_abtem_2021}. Thermal vibration averaging was omitted from the multislice simulation, as the inherent randomness of atomic positions in amorphous materials is larger than such fluctuations. The multislice simulation was set up using a 200 kV probe with a convergence semi-angle of 1.5 mrad and a full width at half maximum of 0.85 \AA. The probe was scanned across the entire simulation box over a $13\times 13$ grid with a step size of 0.51 nm. Each snapshot was sampled using a $2048\times 2048$ grid over the simulation box with a slice thickness of 1 \AA. Partial coherence due to finite source size and beam energy spread were not included.

The resulting time-resolved 4D STEM data has five dimensions: $I(\mathbf{r}, \mathbf{k}, t_w) = I(x,y,k_x,k_y,t_w)$, where $\mathbf{r} = (x,y)$ represents the real space probe position, $\mathbf{k} = (k_x,k_y)$ is the reciprocal space scattering vector, and $t_w$ is the wait time, defined as the time elapsed since the experiment started. Each NBED pattern was polar transformed over the first diffraction ring. The resulting data has the form $I(x,y,k_r,k_\phi,t_w)$, where $k_r$ represents the radius coordinates that range from \SI{3}{nm^{-1}} to \SI{5.5}{nm^{-1}} over 17 pixels, and $k_\phi$ is the azimuthal angle covering the entire $2\pi$ radians of the diffraction ring spread across 180 pixels.

\subsection[Pt57.5Cu14.7Ni5.3P22.5 Nanowire]{Pt\textsubscript{57.5}Cu\textsubscript{14.7}Ni\textsubscript{5.3}P\textsubscript{22.5} Nanowire}

The Pt\textsubscript{57.5}Cu\textsubscript{14.7}Ni\textsubscript{5.3}P\textsubscript{22.5} SCL nanowire data were collected by Huang and Voyles for previous research~\cite{Huang2024,huang2023data}. The ECM experiment was performed using a 1.53 mrad convergence semi-angle probe with a corresponding probe size of 1 nm and a probe current of 25 pA. The probe scanned over a $160\times160$ grid with a step size of 0.28 nm, and an NBED pattern was collected at each scan position with a frame rate of 4000 fps. The scan was continuously repeated to obtain time-resolved 4D STEM data with 250 time steps separated by 6.4 seconds. The time-resolved 4D STEM data was already preprocessed and polar transformed, having the same form of $I(x,y,k_r,k_\phi,t_w)$ as the MD-simulated data. The original data had a $k_r$ range of \SI{3}{nm^{-1}} to \SI{6}{nm^{-1}}, which included some pixels with very low intensity throughout the experiment. Thus, the data was cropped to a final $k_r$ range from \SI{3.56}{nm^{-1}} to \SI{5.44}{nm^{-1}} over 10 pixels.

\subsection{Structural Dynamics Measurements}

\subsubsection{Self-Intermediate Scattering Function}

In glasses and SCLs, structural dynamics are linked to individual atomic movements via the self-intermediate scattering function (self-ISF), which tracks how atoms on average move away from their initial positions at a given experimental wait time~\cite{Janssen2018}. While the self-ISF is not directly measurable in electron scattering experiments, it can be calculated from MD trajectories, where the instantaneous positions of atoms are tracked. The self-ISF, or $f_s$, at a 4D STEM probe position $\mathbf{r}$ can be calculated as~\cite{Scalliet2022}:
\begin{gather}
    f_s(\mathbf{r}, t_w, \Delta t) = \langle\langle \cos[2\pi\mathbf{k_m}\cdot\Delta\mathbf{r_n}(t_w, \Delta t)] \rangle_n\rangle_m, \label{eq:ISF} \\
    \Delta\mathbf{r_n}(t_w,\Delta t) = \mathbf{r_n}(t_w+\Delta t) - \mathbf{r_n}(t_w), \nonumber
\end{gather}
where $\Delta t$ is the delay time, $\mathbf{r_n}$ is the location of the $n$-th atom in the sampling volume at $t_w$, $\mathbf{k_m}$ is a set of reciprocal vectors, and $\langle \dots \rangle_i$ indicates an ensemble average over $i$. To match the real space and reciprocal space sampling of the time-resolved 4D-STEM data under the column approximation, the sampling volume is defined as a cylinder centered at $\mathbf{r}$ with a diameter of 1 nm, and the set of $\mathbf{k_m}$ matches the grid points of the polar-transformed pattern of the first diffraction ring. Conventionally, $f_s$ is defined using a complex exponential form~\cite{Grbel2004}, but for symmetrical $\mathbf{k}$ sampling, the sum over $\pm\mathbf{k}$ reduces the complex exponential function to the real cosine form shown in Equation~\ref{eq:ISF}. For any given wait time, as the delay time increases, the atoms gradually move away from their initial positions, and the atomic structure decorrelates, causing the self-ISF to decrease. The rate at which the self-ISF decays with respect to the delay time represents the local structural dynamics at the moment defined by $t_w$.

\subsubsection{One-Time Autocorrelation}

While the self-ISF cannot be directly measured by ECM, it is linked to the one-time intensity autocorrelation function $g_2$ via the Siegert relation~\cite{Zhang2018,Ragulskaya2024,Grbel2004}, $g_2=Af_s^2$, where $A$ is a constant based on the contrast of the speckles. $g_2$ measures the time-averaged rate of intensity changes of individual speckles throughout the wait time. Since the materials in this research are isotropic, the local $g_2$ function can be obtained by averaging the measurements over all speckles in the first diffraction ring. For each 4D STEM probe position $\mathbf{r}$, the $g_2$ function is~\cite{Fluerasu2005,berne2000dynamic,Livet2001,Lumma2000}
\begin{gather}
    g_2(\mathbf{r},\Delta t) = \langle\{ \delta I(\mathbf{r},\mathbf{k},t_w)\delta I(\mathbf{r},\mathbf{k},t_w+\Delta t) \}_{t_w}\rangle_{\mathbf{k}}, \label{eq:g2} \\
    \delta I(\mathbf{r},\mathbf{k},t_w) = \frac{I(\mathbf{r},\mathbf{k},t_w)-\overline{I}}{\overline{I}},\nonumber
\end{gather}
where $\langle \dots \rangle_{\mathbf{k}}$ is the ensemble average over $\mathbf{k}$, and $\{ \dots \}_{t_w} $ represents the average over the experimental duration. $\delta I$ is the normalized intensity, and $\overline{I}$ is the expected average electron scattering intensity at scattering vector $\mathbf{k}$ and probe position $\mathbf{r}$, serving as the mean reference about which the intensity fluctuates. For most cases, $\overline{I}$ is unknown and must be estimated. Different methods for estimating $\overline{I}$ are discussed and tested in Section~\ref{sec:ESI}.

\subsubsection{Two-Time Autocorrelation}

Another common way to measure the structural decorrelation in XPCS and ECM is through the intensity two-time autocorrelation function $c_2$. Unlike the one-time correlation function $g_2$, $c_2$ measures the ensemble-averaged rate of intensity changes of all speckles starting from a certain wait time. Since no averaging over $t_w$ is required, $c_2$ can capture the time-resolved or "momentary" dynamics of the system~\cite{Cipelletti2002,Fluerasu2005,Bikondoa2017,Ragulskaya2024,Das2020,Das2019,Jain2020,Conrad2015,Riechers2024}. At each 4D STEM probe position $\mathbf{r}$, $c_2$ is
\begin{gather}
    c_2(\mathbf{r}, t_w, \Delta t) = \left\langle \delta I(\mathbf{r}, t_w, \mathbf{k})\delta I(\mathbf{r}, t_w + \Delta t, \mathbf{k}) \right\rangle_\mathbf{k}. \label{eq:c2}
\end{gather}

In XPCS, $c_2$ is commonly represented as a symmetrical two-time correlation function, $TTCF(\mathbf{r}, t_1, t_2) = c_2(\mathbf{r}, t_w=t_1, \Delta t=t_2-t_1)$, to visualize the time-resolved dynamics. The positions along the diagonal of $TTCF$ correspond to different wait times during the experiment, and the width of the diagonal represents the momentary dynamics. $c_2$ can also be used to measure time-averaged dynamics by averaging over the wait time, where $\{ c_2 \}_{t_w}=g_2$. In most previous XPCS and ECM studies, $g_2$ and $c_2$ were defined using different formulas than the ones defined in this section, depending on whether the intensity normalization is explicitly defined in $g_2$ and $c_2$, but the formulas are interchangeable. The relation between the two sets of formulas is discussed in Appendix~\ref{append:acf_def}.

\subsubsection{Structural Relaxation Time Fitting}

$g_2$ and $c_2$ decay with respect to the delay time $\Delta t$ at the same rate as the square of the self-ISF. The decay of these functions typically follows the Kohlrausch-Williams-Watts (KWW) stretched exponential formula, $C\exp\left[-2\left(\frac{\Delta t}{\tau}\right)^{\beta}\right]$, where $\tau$ is the structural relaxation time, $\beta$ is the stretching exponent, and $C$ is a scaling factor to account for experimental partial coherence and initial decorrelation caused by fast relaxation events that are not captured by the time sampling~\cite{Giordano2016}. Three types of structural relaxation parameters are calculated by fitting scale-dependent variants of the self-ISF, $g_2$, and $c_2$ to the KWW form. Global parameters ($\langle\tau\rangle$, $\langle\beta\rangle$) are determined from fully averaged variants ($\langle\{f_{s}\}_{t_{w}}\rangle_{r}$, $\langle g_{2}\rangle_{r}$, $\langle\{c_{2}\}_{t_{w}}\rangle_{r}$); local time-averaged parameters ($\tau(r)$, $\beta(r)$) are obtained by omitting spatial averaging ($\{f_{s}\}_{t_{w}}$, $g_{2}(r,\Delta t)$, $\{c_{2}\}_{t_{w}}$); and local momentary parameters ($\tau_{m}(r,t_{w})$, $\beta_{m}(r,t_{w})$) are extracted from time-resolved forms ($f_{s}(r,t_{w},\Delta t)$, $c_{2}(r,t_{w},\Delta t)$) without temporal averaging. The uncertainties of the relaxation times and stretching exponents were estimated by the standard error of the non-linear least-squares fit. Before fitting the self-ISF and intensity autocorrelation functions using the KWW form, the functions are first logarithmically resampled along the $\Delta t$ axis to balance the weight of the least-squares fit across the relaxing region and the decorrelated tail. The first data point ($\Delta t = 0$) is excluded from the fitting to avoid interference from fast $\beta$-relaxation and experimental noise decorrelation, and the final ten data points are omitted, as they are usually noisy. 

The fitting is performed using Levenberg-Marquardt algorithm~\cite{gavin2019levenberg} via \texttt{scipy.optimize.curve\_fit} function~\cite{2020SciPy-NMeth} with maximum iteration of 1000000. The initial guess of the fitting parameter $C$ in the KWW formula is the value of the first data point ($\Delta t=0$) of the resampled correlation function, and the initial guess of $\beta$ is 1. The initial guess of $\tau$ is set to the value or the estimated global relaxation time of the system. During the fitting process, the fitting parameters are bound to $0.5C_0\leq C\leq 2C_0$, $0.1\leq\beta\leq2$, and $0<\tau$, where $C_0$ is the initial guess of $C$. 

When calculating local relaxation times from the experimental $g_2$ and $c_2$, small averages in real space or the wait time are necessary to reduce noise. For consistency, the following two procedures were applied to both the experimental and simulated datasets. For local $\tau$ and $\tau_m$ calculations, a real space Gaussian weighted averaging of the underlying autocorrelation functions ($g_2$, $c_2$, or the self-ISF) was applied. The $3\times3$ pixel Gaussian kernel has a standard deviation matching the scanning step size. For momentary $\tau_m$ and $\beta_m$ calculations, we applied a moving window average to $c_2$ and the self-ISF along the $t_w$ axis based on the global structural relaxation time $\langle\tau\rangle$ estimated from the self-ISF or $g_2$,
\begin{equation*}
    c_2(t_w, \Delta t) = \frac{1}{n_\tau}\sum_{t_n = t_w-n_\tau/2}^{t_w+n_\tau/2}c_2^\mathrm{ori}(t_n, \Delta t),
\end{equation*} 
where $n_\tau$ is the number of time steps corresponding to one global structural relaxation time.

\subsubsection{Expected Scattering Intensity Estimation} \label{sec:ESI}

For an amorphous material, the statistical expected intensity $\overline{I}=\int [P(I)\times I] dI$, where $P(I)$ is the intensity probability distribution function, is directly linked to the static structure factor $S(k)$. However, because of dynamical scattering, inelastic scattering, and thermal diffuse scattering, it is impractical to compute $\overline{I}$ in an electron scattering experiment even with a known $S(k)$, and $\overline{I}$ must be evaluated from the time series of scattering intensities.

There are two common ways of estimating the expected scattering intensity $\overline{I}$—one for $g_2$ and one for $c_2$—adopted from XPCS. The most straightforward way to estimate $\overline{I}$ is by averaging the scattering intensity at $\mathbf{k}$ over the experiment duration, i.e., $\overline{I}(\mathbf{k})\approx\{ I(\mathbf{k},t_w) \}_{t_w} $ (the $\mathbf{r}$-dependence is omitted hereafter for brevity). Substituting $\overline{I}$ in Equation~\ref{eq:g2} with $\{ I(\mathbf{k},t_w) \}_{t_w} $, we obtain the time-averaged intensity normalized one-time correlation function, which we denote $\tilde{g_2}$. This approach has been used in probe-based~\cite{Huang2024} and DF image-based ECM~\cite{Zhang2018,Chatterjee2021,Vaerst2023} to study time-averaged dynamics in metallic SCLs and glasses.

Another method for estimating $\overline{I}$ is by averaging the intensity over a set of scattering vectors $\mathbf{k}$ at a range of magnitudes and directions, i.e., $\overline{I}(t_w)\approx \langle I(\mathbf{k},t_w) \rangle_{\mathbf{k}} $, assuming every scattering vector shares the same $\overline{I}$ at that specific wait time. Substituting $\overline{I}$ in Equation~\ref{eq:c2} with $\langle I(\mathbf{k},t_w) \rangle_{\mathbf{k}} $ results in the $\mathbf{k}$-averaged intensity normalized two-time correlation function, referred to as $\tilde{c}_2$. This approach has been used in NBED-based ECM to probe time-resolved dynamics~\cite{Huang2024} as well as time-averaged dynamics via $\{ \tilde{c}_2 \}_{t_w}$~\cite{Nakazawa2025,Nakazawa2023Structure}.
 
We propose an alternate, physically-motivated method for estimating $\overline{I}$ in SCLs. Since most SCLs are isotropic, and assuming no major structural changes happen during the experiment, the expected value of the scattering intensity should only be a function of $k_r$. Therefore, the expected scattering intensity can be estimated by the time- and azimuthal averaged intensity $\overline{I}(k_r)\approx\langle \{ I(k_r,k_\phi,t_w) \}_{t_w}  \rangle_{k_\phi}$. Applying this approximation to Equation~\ref{eq:g2} and Equation~\ref{eq:c2} allows the calculation of the time- and azimuthal averaged intensity normalized one-time and two-time correlation functions, which we will refer to as $g_2$ and $c_2$ hereafter.

In experimental data, we noticed that the azimuthal average can sometimes be problematic when estimating $\overline{I}$ using $\langle \{ I(k_r,k_\phi,t_w) \}_{t_w}  \rangle_{k_\phi}$. Because of the diffuse and random nature of the NBED patterns, the residual ellipticity of the first diffraction rings can be difficult to measure and correct perfectly. Small elliptical distortions will cause the pixels corresponding to a $k_r$ coordinate in polar-transformed patterns to contain signals from different radii. To mitigate this effect, instead of an azimuthal average over the entire ring, we employ a moving window average along the $k_\phi$ axis to estimate $\overline{I}$, where $\overline{I}(k_r,k_\phi) = \langle \{ I(k_r,k_\phi,t_w) \}_{t_w}  \rangle_{k_{\phi\pm\Delta\phi}}$. By limiting the azimuthal average to a section of the ring, the variation of $k_r$ within the averaged azimuthal angle range can be minimized. We found that $\Delta\phi=30^{\circ}$ (i.e., a \ang{60} moving window) works well for the experimental data. Because a beam stop was used in the experiment, $\delta I$ calculated from the $\pm \ang{35}$ region from the center of the beam stop is inaccurate, and the data from this region were discarded when calculating $g_2$ and $c_2$.

All of the data processing in this work was parallelized on a high-performance computing cluster using the HyperSpy~\cite{hyperspy2025} and pyXem~\cite{pyxem2025} Python modules. The simulated time-resolved 4D STEM of the CuZr MD model and all of the code and parameters used to calculate and analyze the correlation functions are hosted at the Figshare repository with DOI: 10.6084/m9.figshare.33198822.

\section{Results} \label{Results}

\subsection{MD-Simulated CuZr SCL}

\subsubsection{Global Structural Relaxation}

\begin{figure}[H]
\centering
\includegraphics{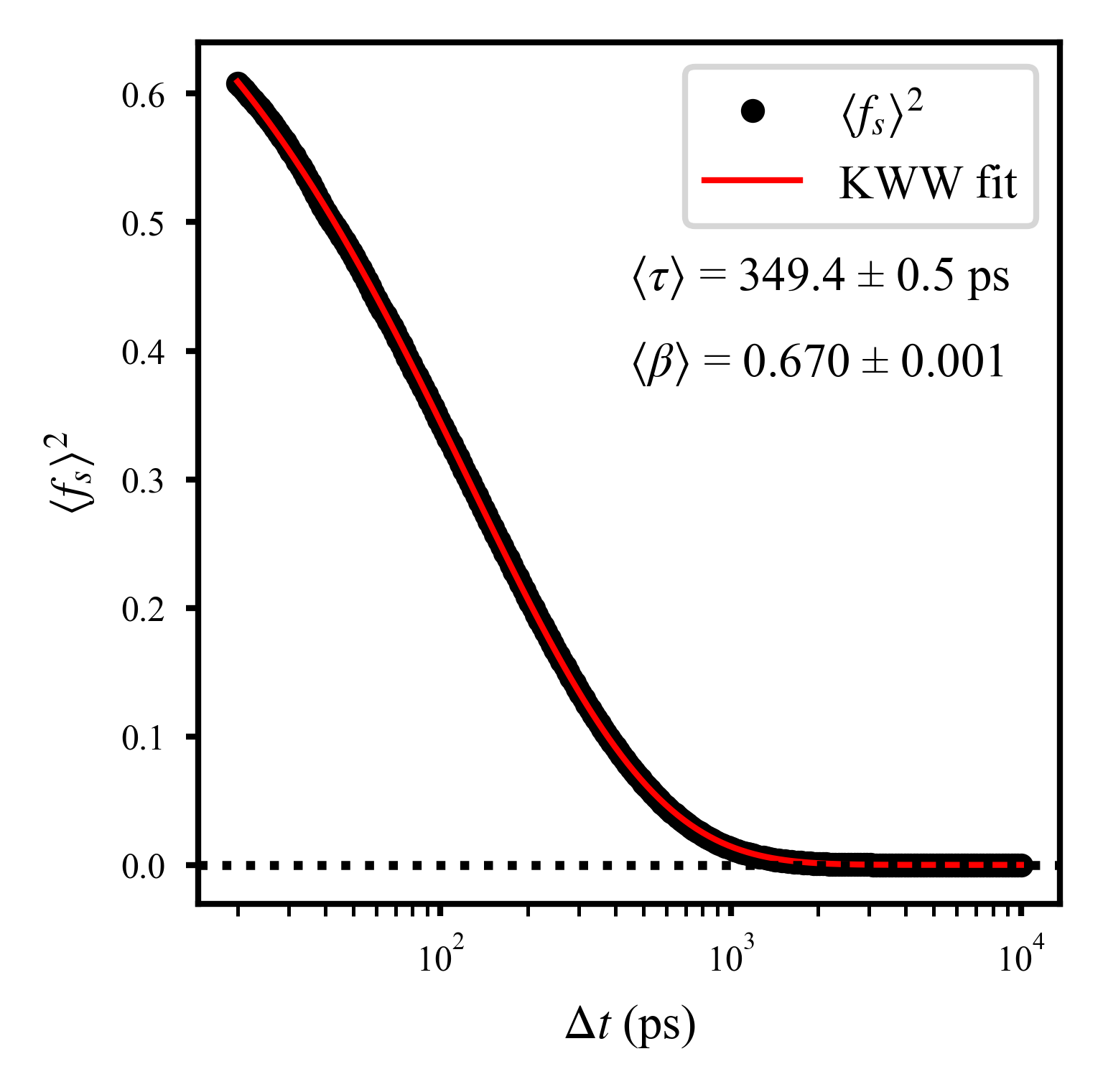}
\caption{Time- and spatially averaged squared self-ISF for the CuZr MD trajectory at 760 K, $T_g+60$ K. The curve was fitted using the KWW function to obtain the global structural relaxation time $\langle\tau\rangle$ and $\langle\beta\rangle$.}
\label{fig:ISF}
\end{figure}

Figure~\ref{fig:ISF} shows the time- and spatially averaged squared self-ISF and the corresponding KWW fit for the CuZr MD trajectory. The self-ISF decay closely follows the KWW form, and the measured global structural relaxation time and stretching exponent are $\langle\tau\rangle=349.4\pm 0.5$ ps and $\langle\beta\rangle=0.670\pm 0.001$, respectively. Using the global structural relaxation time and stretching exponent obtained from the self-ISF as benchmarks, we can compare the accuracies of the structural relaxation time measurements using different intensity autocorrelation functions. 

\begin{figure}[H]
\centering
\includegraphics{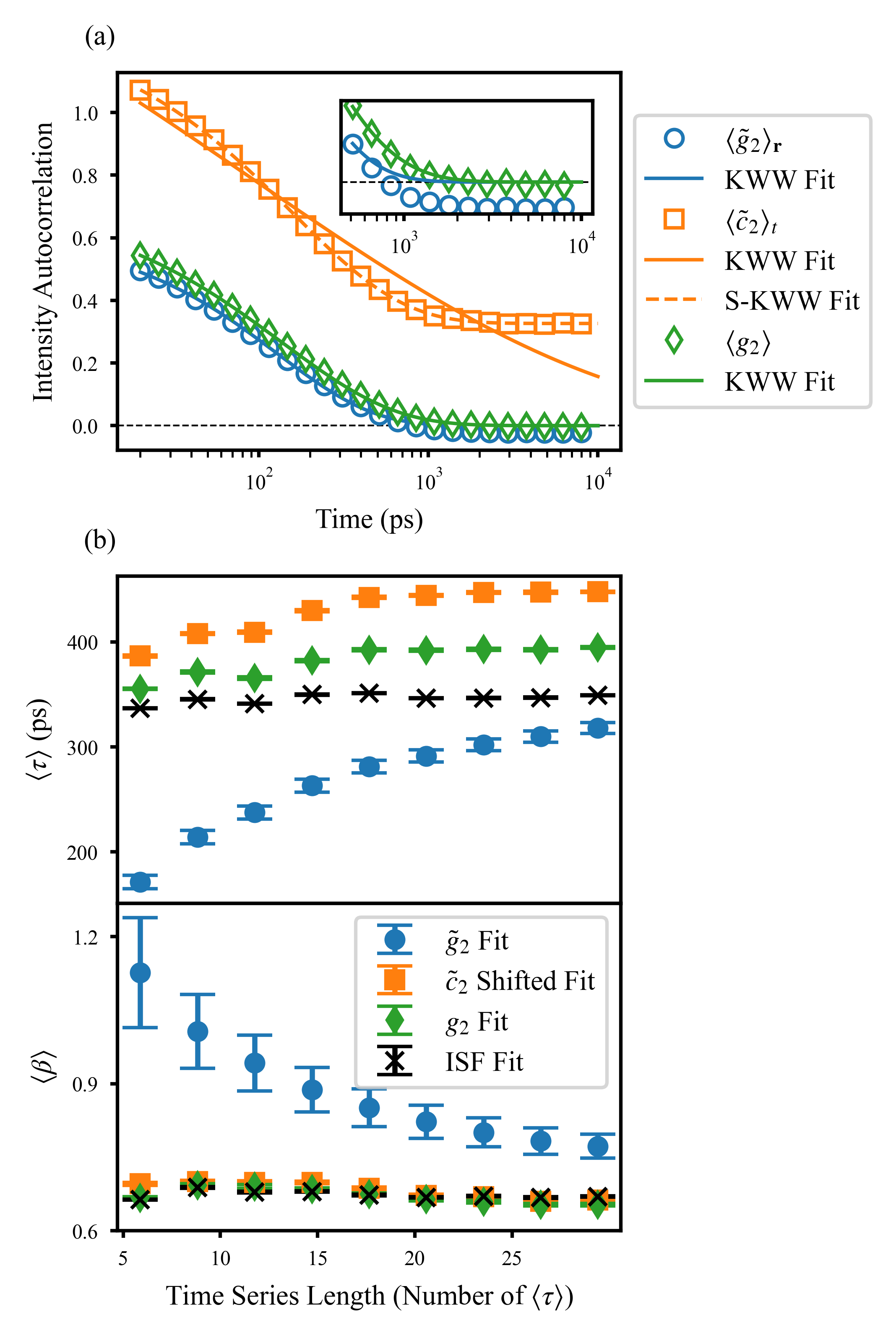}
\caption{(a) Time- and spatially averaged intensity autocorrelation functions calculated based on the three $\overline{I}$ estimation methods: time averaging ($\tilde{g}_2$), $\mathbf{k}$ averaging ($\{ \tilde{c}_2 \}_{t_w}$), and time- and azimuthal ($k_\phi$) averaging ($g_2$). All three autocorrelation functions were fitted with the KWW function, and $\{ \tilde{c}_2 \}_{t_w}$ was additionally fitted with the shifted KWW function. (b) Fitted values (error bars derived from the standard error of the non-linear least-squares fit) of $\langle\tau\rangle$ and $\langle\beta\rangle$ for $\tilde{g}_2$, $\{ \tilde{c}_2 \}_{t_w}$, $g_2$, and the self-ISF. The simulated time-resolved 4D STEM was symmetrically truncated around the temporal midpoint to create subsets with a varying number of time steps. The $\overline{I}$ values, intensity autocorrelation functions, and the self-ISF were calculated from each subset separately to investigate how the temporal duration of the datasets affected the estimated structural dynamics.}
\label{fig:MD_g2_comp}
\end{figure}

Figure~\ref{fig:MD_g2_comp} (a) shows the time- and spatially averaged intensity autocorrelation function normalized by the time-averaged intensity, $\langle\tilde{g}_2\rangle_\mathbf{r}$, the $\mathbf{k}$-averaged intensity, $\langle\{\tilde{c}_2\}_{t_w}\rangle_\mathbf{r}$, and the time- and azimuthal averaged intensity, $\langle g_2\rangle_\mathbf{r}$. At a first glance, $\langle\tilde{g}_2\rangle_\mathbf{r}$ appears to decay similarly to the self-ISF. However, in the high delay time regime shown in the figure inset, $\langle\tilde{g}_2\rangle_\mathbf{r}$ exhibits a negative autocorrelation which is not present in the self-ISF. Since the KWW function cannot drop below 0 by definition, fitting $\langle\tilde{g}_2\rangle_\mathbf{r}$ to the KWW form returns incorrect $\langle\tau\rangle$ and $\langle\beta\rangle$ of $\langle\tau\rangle=318\pm 5$ ps and $\langle\beta\rangle=0.77 \pm 0.02$.

$\langle\{ \tilde{c}_2 \}_{t_w}\rangle_\mathbf{r}$ remains high even at long delay times where the self-ISF has already decayed. This high autocorrelation deviates from the KWW form, resulting in an artificially high relaxation time of $\langle\tau\rangle=2700\pm 900$ ps. Since the shift behaves as a constant background in the autocorrelation function, one way to mitigate this problem is by adding a constant $D$ to the KWW formula as a fitting parameter~\cite{Zhang2023}: $\{ \tilde{c}_2(t_w, \Delta t) \}_{t_w} = C\exp\left[-2\left(\frac{\Delta t}{\langle\tau\rangle}\right)^{\langle\beta\rangle}\right] + D$. Here, the initial guess of $D$ is 0. Fitting $\langle\{\tilde{c}_2\}_{t_w}\rangle_\mathbf{r}$ to this shifted KWW function gives $\langle\tau\rangle=434 \pm 3$ ps and $\langle\beta\rangle=0.584 \pm 0.005$, which are closer to the values extracted from the self-ISF but still inaccurate. However, this method introduces an additional fitting parameter, which can increase the uncertainty of the fit and can erroneously remove actual correlation at the tail region from extremely stable structures.

The time- and azimuthal averaged intensity normalized $\langle g_2\rangle_\mathbf{r}$ most closely matches the self-ISF. It shows no anti-correlation nor abnormally elevated autocorrelation at high delay times. Thus, $g_2$ fit by the KWW functionobtains $\langle\tau\rangle=395.0 \pm 0.9$ ps and $\langle\beta\rangle=0.653\pm 0.002$, which closely match the values from the self-ISF.

Figure~\ref{fig:MD_g2_comp} (b) shows the dependence of $\langle\tau\rangle$ and $\langle\beta\rangle$ on the total duration of the time series for the various correlation functions. For all time series lengths studied, both $\langle\tau\rangle$ and $\langle\beta\rangle$ estimated from $g_2$ closely match the values extracted from the self-ISF, with only a slight overestimation of $\langle\tau\rangle$. $\{ \tilde{c}_2 \}_{t_w}$ is also relatively stable across different time series lengths, with a consistently lower $\langle\beta\rangle$. On the other hand, $\tilde{g}_2$ is much more sensitive to the time series length, returning a much shorter $\langle\tau\rangle$ and a larger $\langle\beta\rangle$ compared with the self-ISF values at short time series lengths. Although the $\langle\tau\rangle$ value gets closer to the self-ISF at long time series lengths, the $\langle\beta\rangle$ value is still higher. Overall, $g_2$ performs the best across different time series lengths for both $\langle\tau\rangle$ and $\langle\beta\rangle$.

Among the three autocorrelation functions, the proposed time- and azimuthal averaged intensity normalized $g_2$ performs demonstrably better than the other two functions. The $g_2$ curve matches the self-ISF the closest, does not require additional fitting parameters, and consistently estimates $\langle\tau\rangle$ and $\langle\beta\rangle$ close to the self-ISF measured values regardless of the simulation time series length. Therefore, from this point forward, we will use $g_2$ and the corresponding $c_2$ to estimate the structural dynamics of SCLs.

\subsection{Spatially Resolved, Time-Averaged Dynamics}

\begin{figure}[H]
\centering
\makebox[\textwidth][c]{\includegraphics{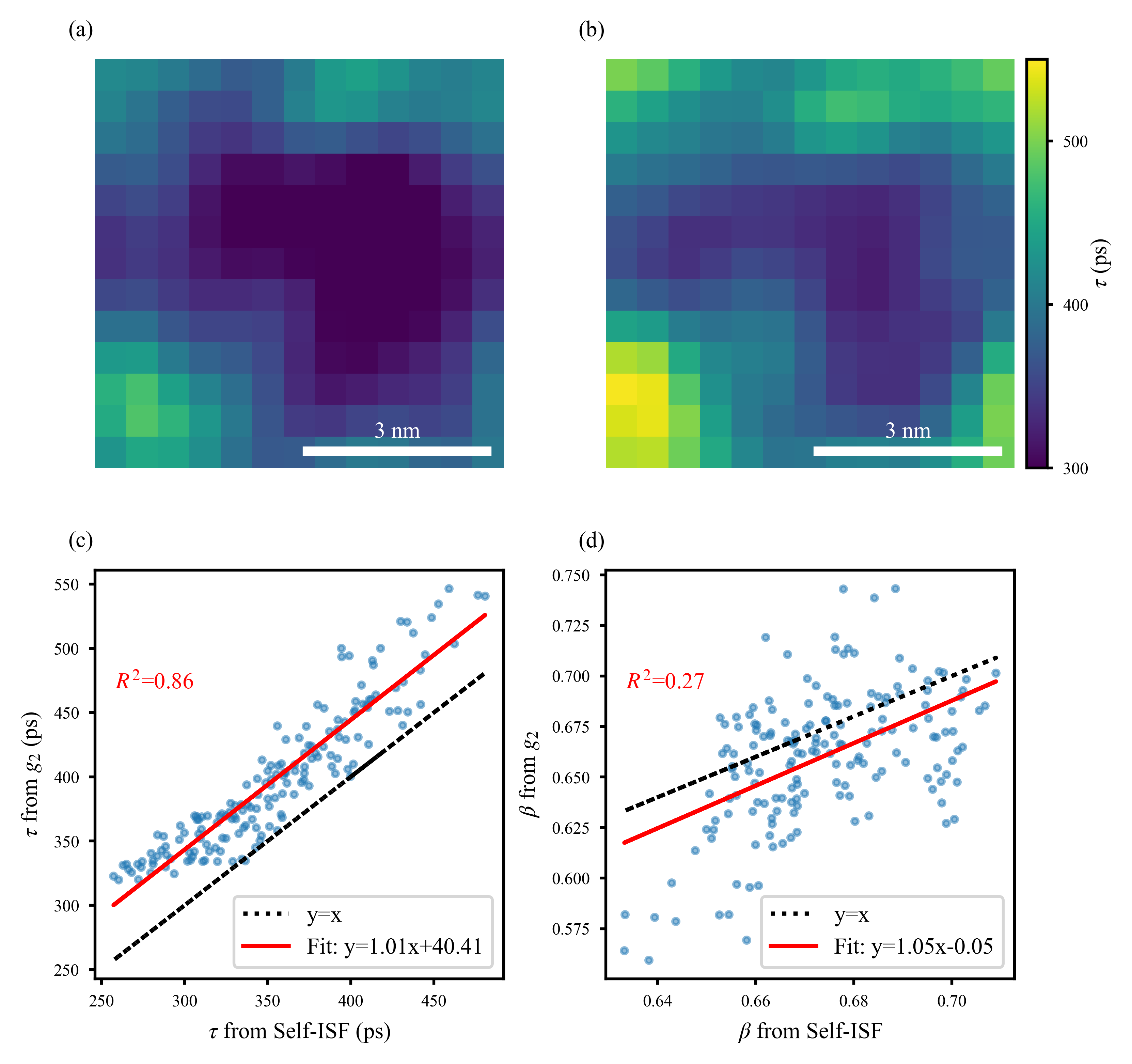}}
\caption{(a,b) Time-averaged structural relaxation time $\tau$ map calculated from the self-ISF and $g_2$, respectively; (c) $\tau$ and (d) $\beta$ from $g_2$ vs. from the self-ISF at each probe position. The red line indicates the regression line of the two sets of $\tau$ and $\beta$ and the black line is equality of the two. The two sets of time-averaged $\tau$ are highly correlated, with the $g_2$ function returning roughly 10 \% longer relaxation times across most probe positions. The correlation between the two sets of $\beta$ is weaker, but the absolute values from both functions are with 0.55 and 0.75, suggesting little $\beta$ variation across the simulation box.}
\label{fig:MD_tau_map}
\end{figure}

While the global dynamics can be measured via the spatially averaged $g_2$ with proper estimation of $\overline{I}$, the primary advantage of NBED-ECM lies in its ability to resolve dynamic heterogeneity at the nanometer scale. Here we examine how the local $g_2$ can capture the spatially resolved dynamics of our model SCL. Figure~\ref{fig:MD_tau_map} (a) and (b) respectively show the distribution of the local structural relaxation time calculated from the self-ISF and $g_2$ at each probe position across the simulation box, illustrating the time-averaged spatially heterogeneous dynamics of the SCL. The local $\tau$ and $\beta$ estimated from $g_2$ and from the self-ISF at each probe position are shown in Figure~\ref{fig:MD_tau_map} (c) and (d), respectively. The structural relaxation times extracted from the two functions are highly correlated, with a slight overestimation for most probe positions from $g_2$, which is consistent with the trend observed in the spatially and time-averaged dynamics. Although the stretching exponents are not strongly correlated, the $\beta$ values vary by less than 0.2, indicating minimal spatial variation across the simulation box. Overall, the local $g_2$ provides a good estimate of the local time-averaged dynamics.

\subsection{Spatially and Time-Resolved Dynamics}

\begin{figure}[H]
\centering
\makebox[\textwidth][c]{\includegraphics{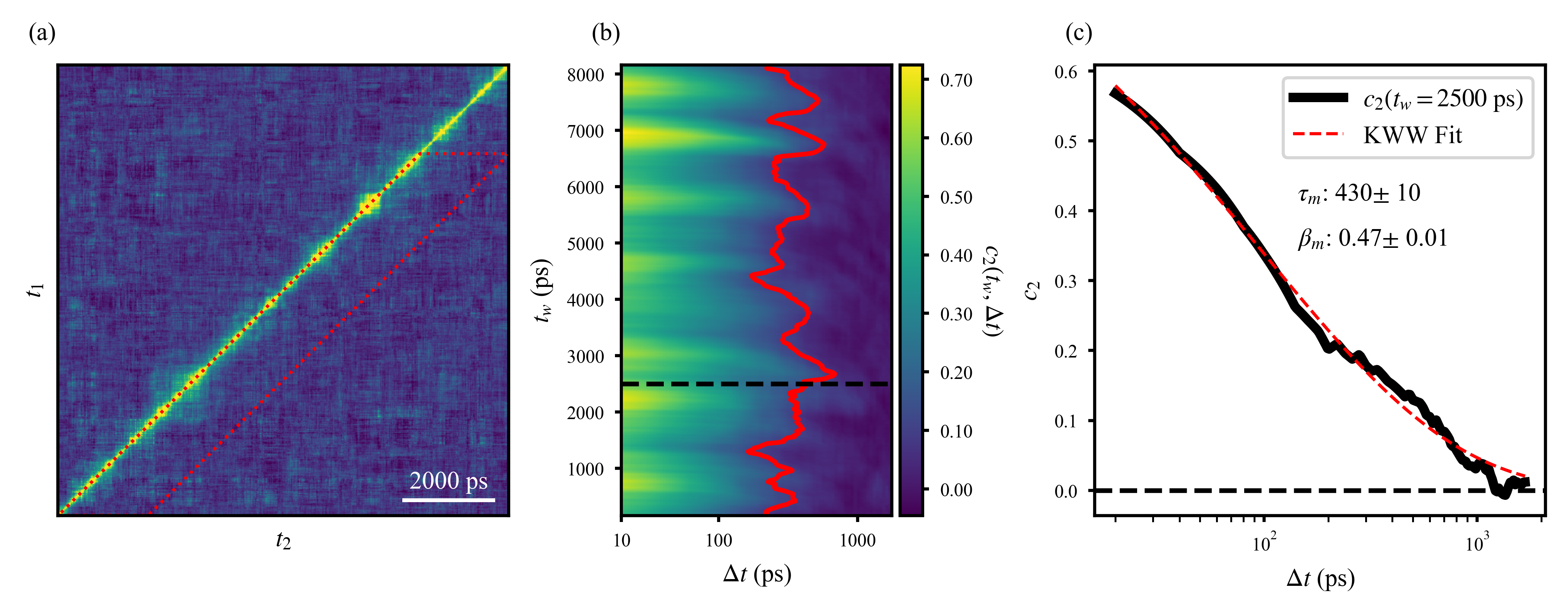}}
\caption{(a) Single probe position TTCF from the center of the MD simulation box; (b) $c_2$ from the red parallelogram marked in (a), where $t_w=t_1$ and $\Delta t=t_2-t_1$. The $c_2$ function is logarithmically resampled along $\Delta t$. The red line marks the momentary relaxation time $\tau_m$ at any given $t_w$; (c) $c_2$ for $t_w=2500$ ps along the black dashed line in (b). $c_2$ at each $t_w$ can be fitted to the KWW function to estimate the momentary structural relaxation time $\tau_m$ and $\beta_m$.}
\label{fig:MD_ttcf_center} 
\end{figure}

Figure~\ref{fig:MD_ttcf_center} (a) shows the single probe position TTCF from the center of the model SCL, where small high correlation patches (slow dynamics) are distributed along the diagonal, suggesting time variation in the dynamics. To quantify the momentary dynamics, the red parallelogram region of the TTCF was extracted as a rectangular matrix of $c_2(t_w,\Delta t)$, as shown in Figure~\ref{fig:MD_ttcf_center} (b). Each row of the matrix represents $c_2$ with a certain wait time $t_w$, which captures the rate the diffraction pattern decorrelates from the initial pattern at $t_w$ as the delay time increases. Figure~\ref{fig:MD_ttcf_center} (c) shows the $c_2$ at $t_w=2500$ ps, corresponding to the black dashed line in Figure~\ref{fig:MD_ttcf_center} (b). For each $t_w$, $c_2$ can be fitted to the KWW form to estimate the momentary structural relaxation time $\tau_m$ and the momentary stretching exponent $\beta_m$~\cite{Das2019,Das2020,Riechers2024}. 

\begin{figure}[H]
\centering
\makebox[\textwidth][c]{\includegraphics{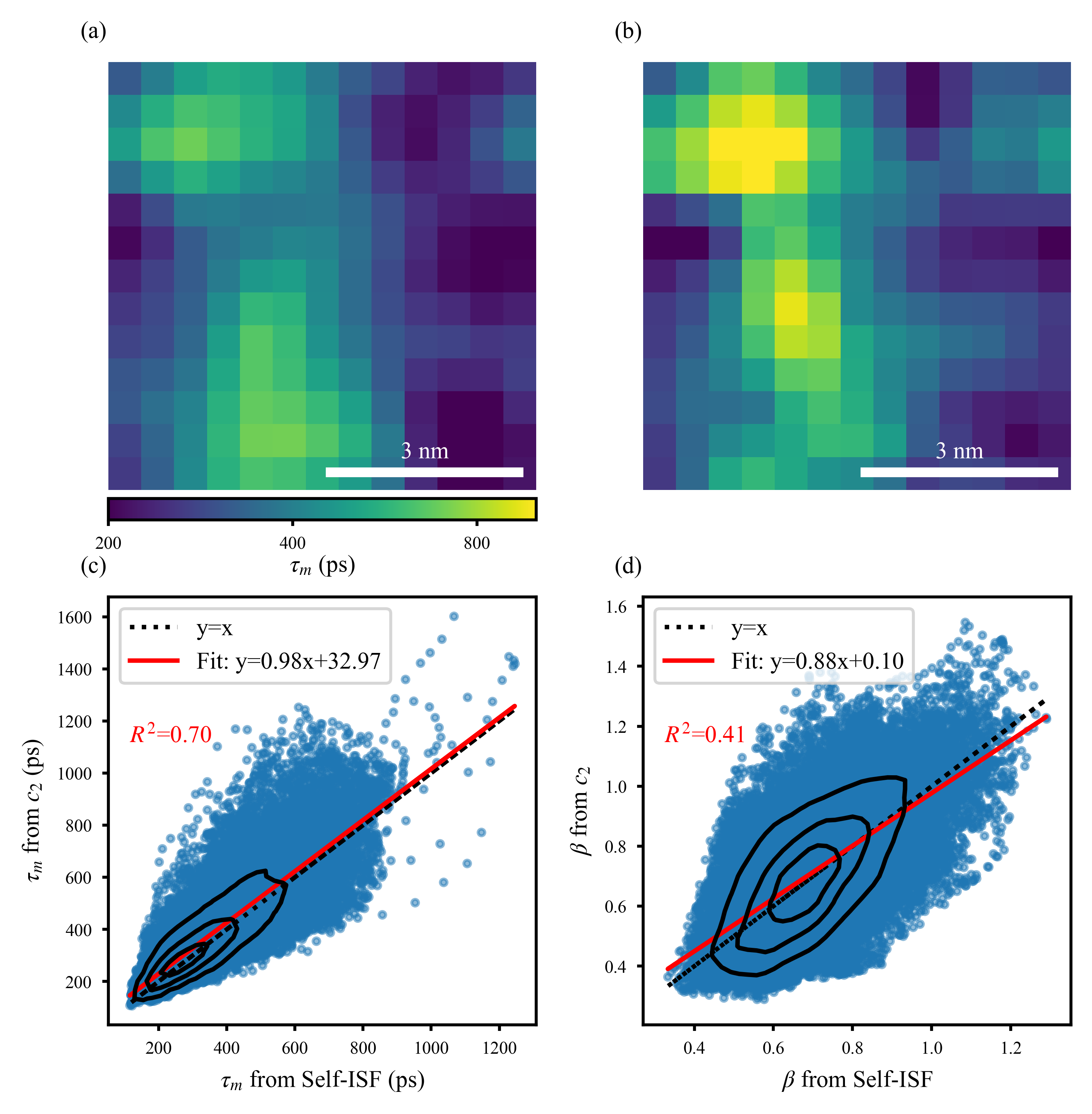}}
\caption{(a,b) Momentary structural relaxation time $\tau_m$ map calculated from the self-ISF and $c_2$, respectively, at $t_w=2500$ ps; (c,d) $\tau_m$ and $\beta_m$ from $c_2$ vs. from the self-ISF at each probe position and wait times, respectively. The red line indicates the regression line and the black line is equality. The contour lines represent density iso-proportions at 20\%, 50\%, and 80\% via kernel density estimation, bounding regions of increasing data concentration toward the distribution center. The two sets of $\tau_m$ and $\beta_m$ are largely correlated, but the $\tau_m$ correlation is weaker compared with the time-averaged dynamics shown in Figure~\ref{fig:MD_tau_map} (c).}
\label{fig:MD_mtau_map}
\end{figure}

Figure~\ref{fig:MD_mtau_map} (a) and (b) show the $\tau_m$ maps calculated at $t_w=2500$ ps from the self-ISF and TTCF, respectively. The two maps demonstrate the same pattern of $\tau_m$ at different positions, but the quantitative agreement is noticeably worse compared to the time-averaged $\tau$ maps shown in Figure~\ref{fig:MD_tau_map} (a) and (b). Figure~\ref{fig:MD_mtau_map} (c) and (d) compares $\tau_m$ and $\beta_m$ from $c_2$ and the self-ISF at all probe positions and wait times, respectively. While the results from both functions correlate positively, they show greater dispersion than the time-averaged $\tau$ and $\beta$ metrics owing to extreme local values of $\tau_m$ and $\beta_m$. Nevertheless, kernel density estimation of the distribution of the calculated $\tau_m$ and $\beta_m$ values shows that the majority of the data points are concentrated near the equality line, indicating a strong correlation between both functions, as demonstrated by the density iso-proportion contour lines in Figure~\ref{fig:MD_mtau_map} (c) and (d).

\subsection[Structural Dynamics of a Pt57.5Cu14.7Ni5.3P22.5 SCL Nanowire]{Structural Dynamics of a Pt\textsubscript{57.5}Cu\textsubscript{14.7}Ni\textsubscript{5.3}P\textsubscript{22.5} SCL Nanowire}

\begin{figure}[H]
\centering
\makebox[\textwidth][c]{\includegraphics{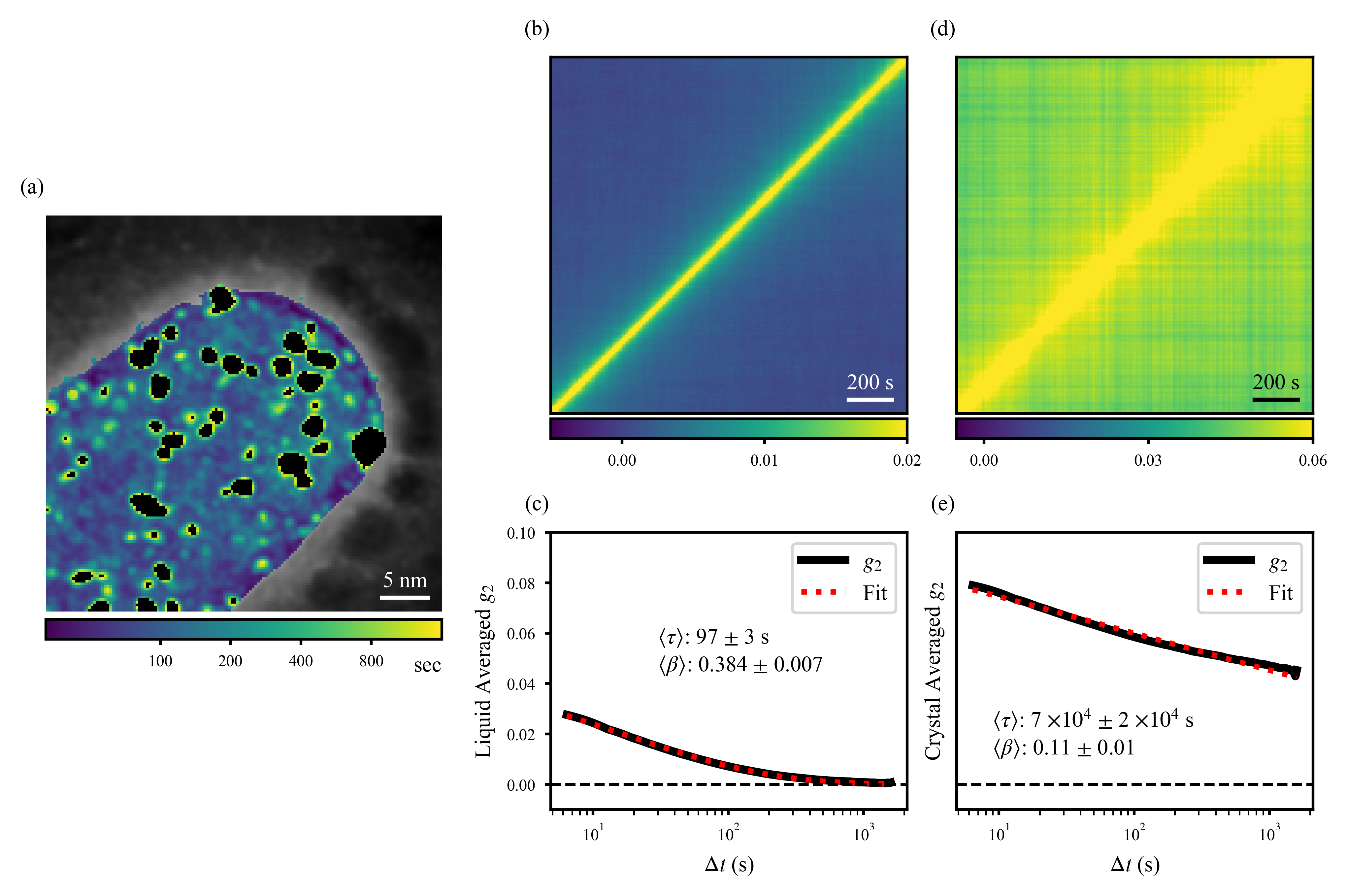}}
\caption{(a) Time-averaged $\tau$ map of the Pt\textsubscript{57.5}Cu\textsubscript{14.7}Ni\textsubscript{5.3}P\textsubscript{22.5} nanowire overlaid on the virtual dark-field image of the nanowire. Regions with relaxation times larger than 1600 seconds are deemed as crystal-like domains and are marked in black; (b,c) spatially averaged TTCF and $g_2$ of the non-crystal regions; (d,e) spatially averaged TTCF and $g_2$ of the crystal-like regions.}
\label{fig:Pt_avg}
\end{figure}

The new NBED-ECM analysis was applied to a time-resolved 4D STEM dataset for a Pt\textsubscript{57.5}Cu\textsubscript{14.7}Ni\textsubscript{5.3}P\textsubscript{22.5} SCL nanowire collected in a previous study at $T_g+4$ K~\cite{Huang2024}. Regions of the wire were crystallized during the ECM experiment and no longer change with time, so they interfere with the calculation of the averaged liquid relaxation time from simply averaging $g_2$ over the entire nanowire. To handle the crystals, the spatially resolved $\tau$ map was first calculated for each probe position using the local $g_2$. Figure~\ref{fig:Pt_avg} (a) shows the distribution of the time-averaged $\tau$ within the nanowire. Multiple regions in the nanowire have an estimated relaxation time larger than 1600 seconds, which is the entire duration of the experiment, showing that the structure of these domains never decorrelated during the experiment. These domains are deemed to be crystal-like domains, and are marked in black in Figure~\ref{fig:Pt_avg} (a). Once the crystal-like regions are identified, the averaged dynamics of the liquid can be measured, as shown in Figure~\ref{fig:Pt_avg} (b) and (c). The spatially averaged TTCF shows that within the experiment duration, the liquid region was in equilibrium, and the averaged $g_2$ estimates a global structural relaxation time of $\langle\tau\rangle=97\pm 3$ seconds with a stretching exponent of $\langle\beta\rangle=0.384 \pm 0.007$. The TTCF and $g_2$ of the crystal-like regions are shown in Figure~\ref{fig:Pt_avg} (d) and (e), where the intensity autocorrelation functions remain high throughout the delay time, confirming the structural stabilities of these domains. 

\begin{figure}[H]
\centering
\makebox[\textwidth][c]{\includegraphics{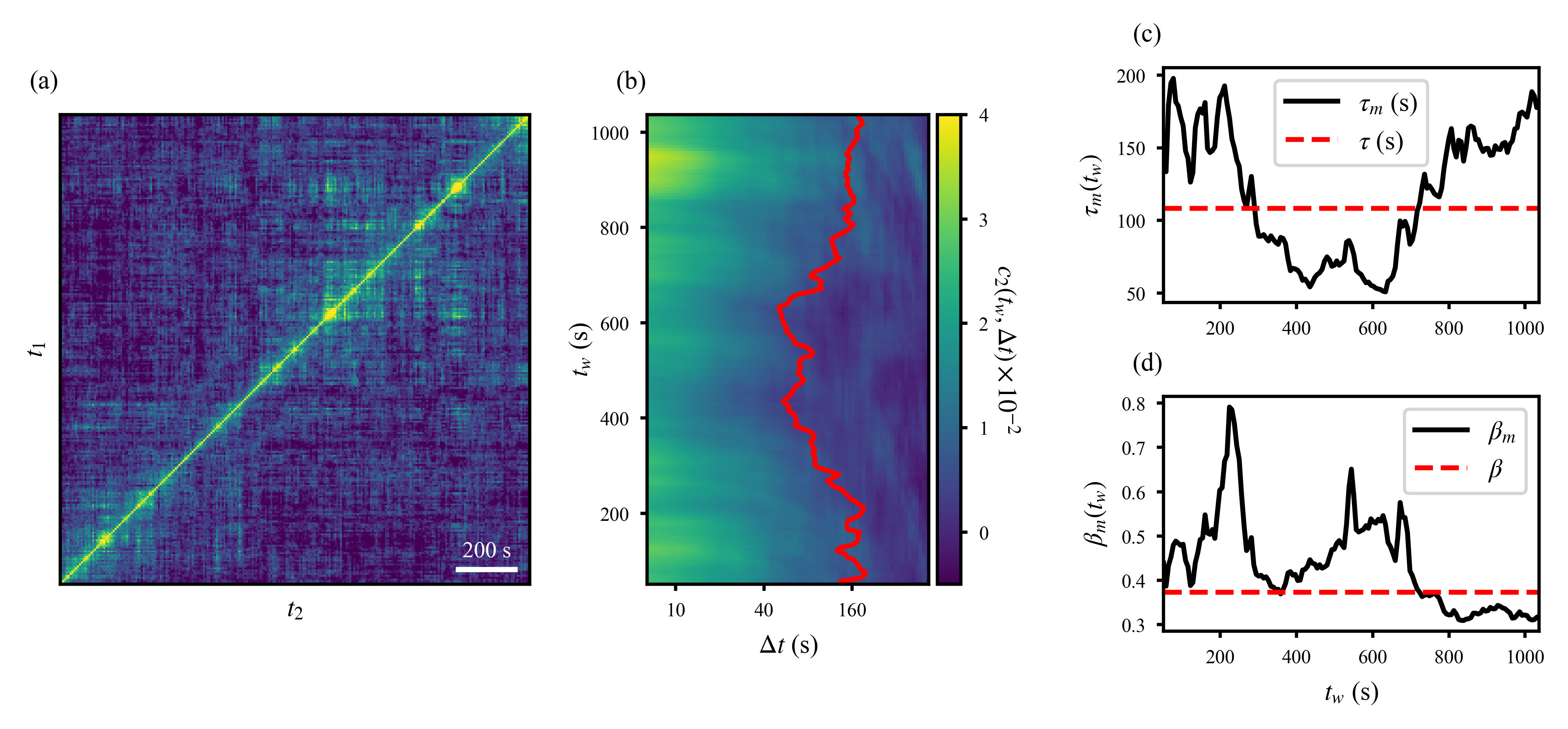}.png}
\caption{(a,b) Local TTCF from the center of the nanowire and the corresponding $c_2$, respectively; (c,d) $\tau_m$ and $\beta_m$ calculated from the $c_2$ in (b). The red line in (b) marks the $\tau_m$ at each $t_w$, and the red dashed line in (c,d) indicates the time-averaged $\tau$ and $\beta$ calculated from the local $g_2$.}
\label{fig:Pt_ttcf}
\end{figure}

Figure~\ref{fig:Pt_ttcf} shows the local TTCF, $c_2$, and the $\tau_m$ and $\beta_m$ from a liquid region in the center of the nanowire. Similar to the MD modeled CuZr SCL, high correlation patches can be found along the diagonal of the TTCF and the momentary relaxation time fluctuates throughout the experiment duration, with the longest $\tau_m$ roughly four times larger than the shortest. These results indicate that although the liquid region of the nanowire was in equilibrium, it still exhibited intermittent, temporally heterogeneous dynamics at a single spatial position.

\begin{figure}[H]
\centering
\makebox[\textwidth][c]{\includegraphics{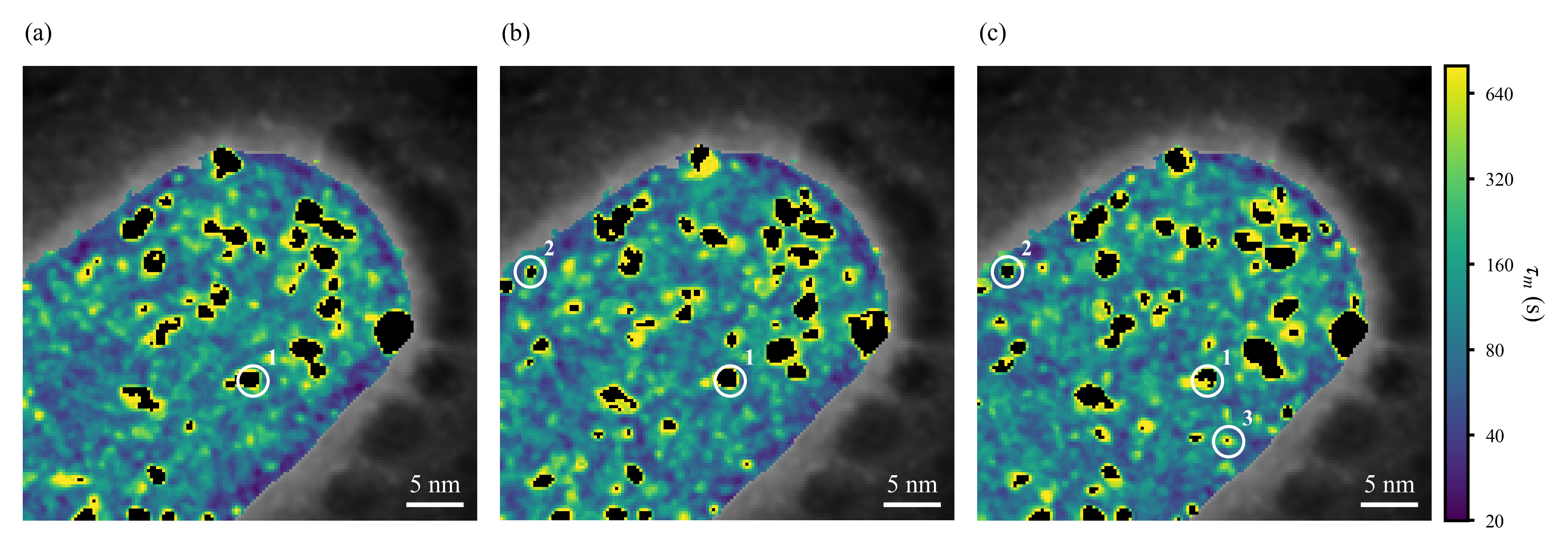}}
\caption{Real space map of the momentary relaxation time $\tau_m$ of the nanowire at three different wait times. The crystal-like regions with $\tau_m > 1600$ seconds are marked as black. Three domains that became crystal-like at different moments of the experiment are marked by the white circles.}
\label{fig:Pt_mtau}
\end{figure}

The momentary relaxation time $\tau_m$ maps of the nanowire at three different wait times are shown in Figure~\ref{fig:Pt_mtau}. The domains with $\tau_m$ larger than 1600 seconds are deemed as crystal-like and marked as black. Most of the crystal-like domains identified in the time-averaged $\tau$ map also appear as crystal-like regions in the time-resolved $\tau_m$ maps. One such domain is marked and labeled as crystal 1 in all three $\tau_m$ maps. However, there are some domains that only appear in the $\tau_m$ maps at later wait times, suggesting that new crystal-like domains formed during the experiment. For example, domain 2 only appears as a crystal in the maps at wait times of 640 seconds and 960 seconds, and domain 3 only at 960 seconds. 

\begin{figure}[H]
\centering
ax\makebox[\textwidth][c]{\includegraphics{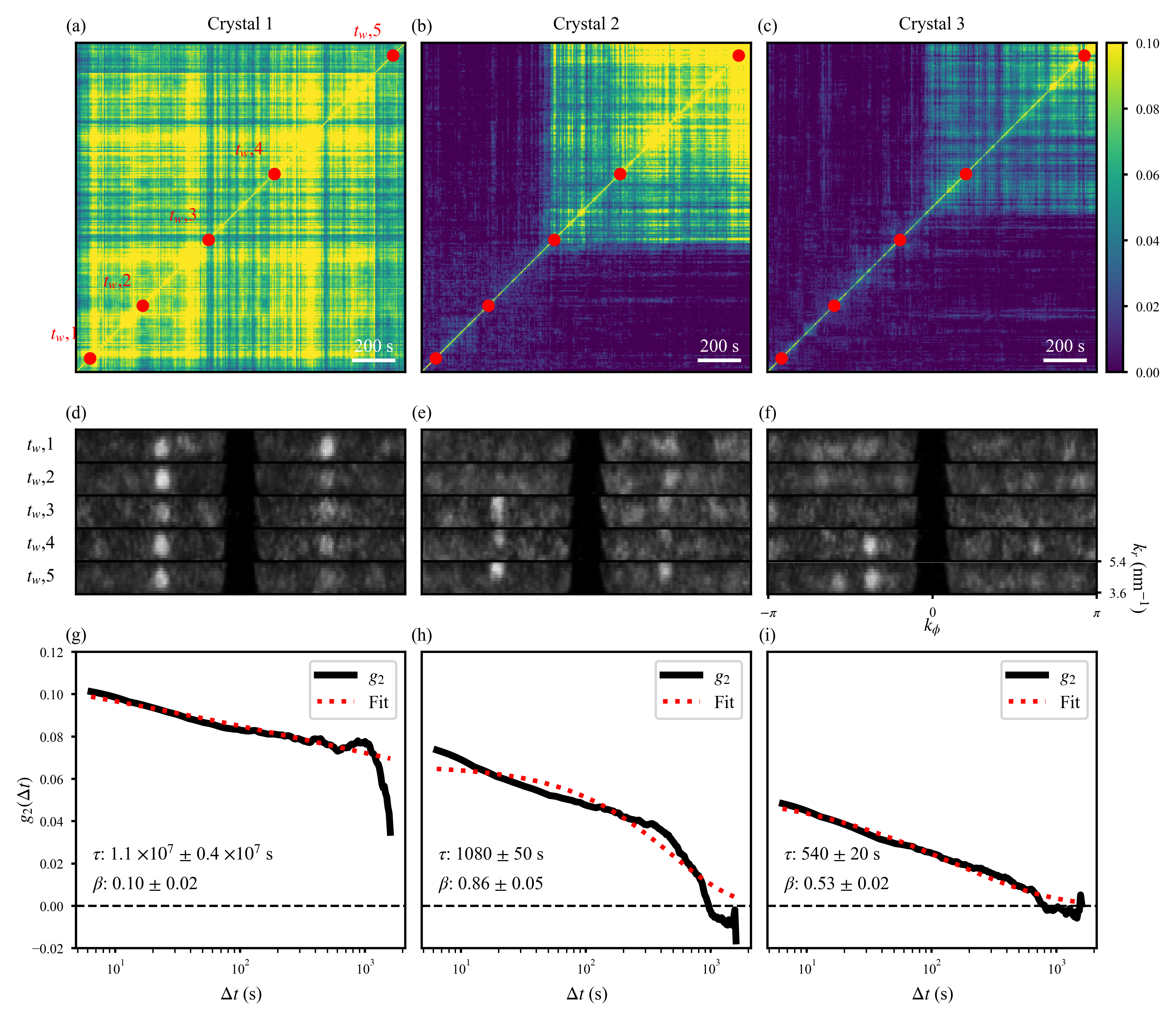}}
\caption{(a-c) Local TTCFs of the domains marked by the white circles labeled 1 to 3 in Figure~\ref{fig:Pt_mtau}. Five moments during the experiment were marked on the TTCFs, where $t_{w,1}$ and $t_{w,5}$ are near the beginning and end of the experiment, respectively, and $t_{w,2}$ to $t_{w,4}$ correspond to the three wait times used to calculate the $\tau_m$ maps shown in Figure~\ref{fig:Pt_mtau}; (d-f) polar-transformed first diffraction ring patterns from the three domains at the five wait times marked in (a-c); (g-i) local $g_2$ of the three domains and the corresponding KWW fit.}
\label{fig:Pt_crystals}
\end{figure}

The local TTCFs from the crystal 1, 2, and 3 positions are shown in Figure~\ref{fig:Pt_crystals} (a-c), where the moment of crystallization can be identified. For crystal 1, the crystallization happened before the experiment, and the structure never decorrelated throughout the wait time. For crystal 2 and 3, the liquid structure decorrelated rapidly at the beginning, and then stabilized upon crystallization happened. The diffraction patterns from the three domains at five wait times are displayed in Figure~\ref{fig:Pt_crystals} (d-f). The five wait times are marked along the diagonal of the TTCFs, where $t_{w,1}$ and $t_{w,5}$ are near the beginning and end of the experiment, and $t_{w,2}$ to $t_{w,4}$ correspond to the three wait times used to construct the $\tau_m$ maps shown in Figure~\ref{fig:Pt_mtau}. Crystal 1 shows strong crystal diffraction spots from the beginning till the end of the experiment. For crystal 2 and 3, the diffraction patterns show typical liquid speckles at the beginning, and crystal peaks appear during the experiment. The wait times that crystal peaks appeared match the wait times when high correlation values are measured in TTCFs, demonstrating the ability of the TTCF to detect these crystallization events. The local $g_2$ from the three crystal positions are shown in Figure~\ref{fig:Pt_crystals} (g-i). All three positions show higher autocorrelation than the averaged liquid $g_2$ and extended delay time due to the crystal formation. However, because $g_2$ also captures the liquid behaviors before the crystal formation, the time-averaged relaxation times for crystal 2 and 3 are shorter than 1600 seconds, and they are not marked as crystal in the time-averaged $\tau$ map. 

\section{Discussion}

\subsection{Effect of Spatial and Temporal Averaging in self-ISF vs ECM Relaxation Time Measurements}

Overall, simulated ECM captures local dynamics consistent with the self-ISF when using $g_2$ and $c_2$. With less averaging, intermittent, temporally heterogeneous dynamics are observed. Spatial heterogeneity is similar in domain patterns in ECM and the self-ISF, as shown in Figure~\ref{fig:MD_tau_map}. 
However, we observed an increase of disagreement between the structural dynamics calculated from the self-ISF and intensity autocorrelation functions ($g_2$ and $c_2$) as the degree of spatial and temporal averaging decreases, as shown in Figure~\ref{fig:MD_mtau_map}. We believe this disagreement is intrinsic to the ECM autocorrelation functions. Unlike the self-ISF, which tracks all atomic movements, ECM tracks the changes only of atoms that contribute to strong electron diffraction. These atoms are a subset of atoms within the interaction volume, because diffraction is sensitive to the orientations of the atomic structures with respect to the electron beam direction~\cite{Hirata2014}. Consequently, some atomic motions that contribute to the self-ISF may be invisible to ECM. However, which atoms contribute to the ECM signal is random. Thus, by sampling over a large number of atoms, the fluctuations inherent to the random subsets are reduced and the ECM vs. self-ISF agreement improves.

\begin{figure}[H]
\centering
\includegraphics{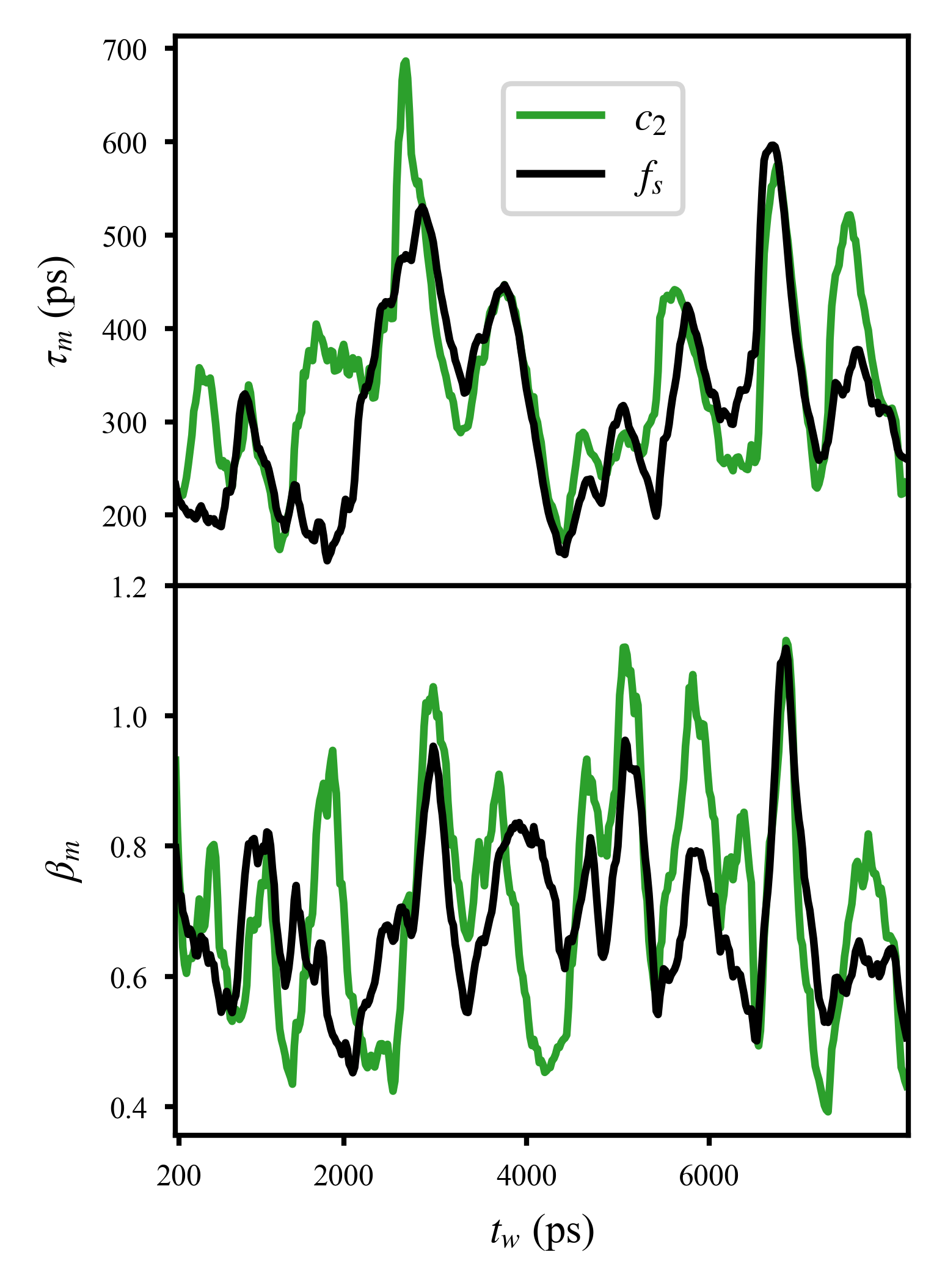}
\caption{$\tau_m$ and $\beta_m$ at each $t_w$ from the center of the MD simulation box, calculated from $c_2$ and the self-ISF. While the values of $\tau_m$ and $\beta_m$ from $c_2$ do not perfectly match the ones from the self-ISF, they are strongly correlated for most $t_w$. However, there are still moments (e.g., $t_w\leq 2000$ ps) where the momentary dynamics estimated by $c_2$ do not reflect the fluctuations shown in the self-ISF curves.}
\label{fig:MD_mtau_mbeta}
\end{figure}

Although the subsets of atoms ECM samples differ from the self-ISF, any change in electron scattering signals is still linked to atomic movements, and thus the time-resolved dynamics estimated from $c_2$ still reflect a subset of the actual dynamics, with $R^2=0.70$ for the linear regression of $\tau_m$ from $c_2$ vs from the self-ISF. The link between local $c_2$ and local self-ISF is demonstrated by Figure~\ref{fig:MD_mtau_mbeta}, which shows the momentary relaxation time and momentary stretching exponent for each wait time calculated from the $c_2$ shown in Figure~\ref{fig:MD_ttcf_center} (b) and from the local self-ISF at the same probe position. While the values are not perfectly matched, the $\tau_m$ and $\beta_m$ from $c_2$ do largely follow the fluctuations observed in the $\tau_m$ and $\beta_m$ from the self-ISF.

\subsection{Systematic Errors in Intensity Correlation Functions}

Figure~\ref{fig:MD_g2_comp}(a) illustrates the systematic errors arising from improper intensity normalization. Specifically, the time-averaged, intensity-normalized $\tilde{g}_2$ function introduces a negative correlation, whereas the $\mathbf{k}$-averaged, intensity-normalized $\tilde{c}_2$ function introduces an elevated baseline correlation at long delay times. For $\tilde{g}_2$, the issue stems from trying to estimate the expected value of a temporally varying signal using its own time average. This estimation reduces the degrees of freedom of the signal and forces an additional constraint, $\{\delta I\}_{t_w}=0$. Mathematically, it can be shown that autocorrelation functions defined this way will always become negative at high delay times, appearing as an artificial anti-correlation~\cite{Percival1993,Nakazawa2025}, as shown in Figure~\ref{fig:MD_g2_comp} (a). Let the intensity time series consist of $N$ discrete frames recorded at times $t_n=n\delta t$, where $n$ is the frame index and $\delta t$ is the experimental sampling interval (frame time). The physical delay time is expressed as $\Delta t=m\delta t$, where $m$ is the dimensionless integer frame delay. There are two common ways to calculate the time-averaged intensity, a uniform average:
\begin{equation}
   \{ I(t_w) \}_{t_w}=\{ I(t_w+\Delta t) \}_{t_w} = \frac{1}{N}\sum_{n=1}^{N} I(t_n), \nonumber
\end{equation}
or a split average~\cite{Zhang2017,He2015,Vaerst2023,Chatterjee2021,Huang2024}:
\begin{equation*}
    \{ I(t_w) \}_{t_w} = \frac{1}{N-m}\sum_{n=1}^{N-m} I(t_n) \ \text{and} \ \{ I(t_w+\Delta t) \}_{t_w} = \frac{1}{N-m}\sum_{n=m}^{N} I(t_n).
\end{equation*}
The averaged $\tilde{g}_2$ curve shown in Figure~\ref{fig:MD_g2_comp} (a) was calculated using the uniform averaging, which produces averaged $\tilde{g}_2$ curves that are negative throughout the tail region. When split averaging is used, $\tilde{g}_2$ will still become negative before being forced to converge to 0 at the very end of the curve. Especially on a $\log$-$\Delta t$ scale, the negative autocorrelation region often appears as a small dip right before the curve reaches 0. Examples of such behavior can be seen in the $\tilde{g}_2$ curves shown in some previous studies~\cite{Vaerst2023,Nakazawa2025,Lumma2000}. At a glance, the $\tilde{g}_2$ curves calculated from split averaging might appear to behave more properly, as they eventually converge to 0. However, the deviation from the KWW form starts when the curves first become negative, and the estimation of $\tau$ and $\beta$ is still affected. Both time averaging methods produce almost identical $\tau$ and $\beta$ estimations in our MD simulated CuZr SCL.

The $\mathbf{k}$-averaged intensity normalized $\tilde{c}_2$ remains too high instead of converging to 0 at long delay times. This additional correlation is due to the failure of the assumption that $\overline{I}$ is the same for all sampled $\mathbf{k}$ vectors. In NBED patterns, because of the convergent beam, the speckles in the first diffraction ring have a finite size that spans across a range of $k_r$, in which the expected intensities vary non-trivially due to the change of the structure factor. On average, the center of the diffraction ring has higher intensity than the inner and outer edges. Thus, $\delta\tilde{I}(\mathbf{k},t_w)$ calculated using $\langle I \rangle_{\mathbf{k}}$ over the whole ring in $k_r$ is almost always positive for $\mathbf{k}$ near the center of the ring, and always negative for those near the edges, artificially creating an extra correlation throughout the whole $\Delta t$ range. This extra correlation can be considered as a constant shift of the autocorrelation function by approximately the normalized variance of the time-averaged diffraction pattern (see Appendix~\ref{append:c2}). This constant shift disagrees with the KWW form of the correlation function, and a regular KWW fit will return completely wrong $\tau$ and $\beta$.

\subsection{Impact of Observation Time and the Pitfalls of Short-Duration Extrapolation}

While conventional normalization frameworks introduce systematic artifacts, the practical severity of these errors heavily depends on the duration of the experiment relative to the system's relaxation time. Because the anti-correlation in $\tilde{g}_2$ (i.e., $\overline{I}=\{ I \}_{t_w}$) is a consequence of reduced degrees of freedom in the temporal signal, the effect of this error can be mitigated by increasing the length of the data time series. The longer the time series, the smaller the impact of reducing one degree of freedom due to the mean estimation. This effect is method-agnostic and will affect the $\tilde{g}_2$ from both tilted-DF and NBED ECM, as well as XPCS, as long as the expected intensity is estimated only with time averaging. Previously, Zhang et al.~\cite{Zhang2017} measured that the required experiment duration for the estimated $\langle\tau\rangle$ from $\tilde{g}_2$ to converge is $40\langle\tau\rangle$ in two metallic SCLs (Pt\textsubscript{57.5}Cu\textsubscript{14.7}Ni\textsubscript{5.3}P\textsubscript{22.5} and Pd\textsubscript{40}Ni\textsubscript{40}P\textsubscript{20}) using tilted-DF ECM. Achieving the $40\langle\tau\rangle$ limit often requires prolonged experimental durations. Consequently, measuring dynamics accurately using $\tilde{g}_2$ is less practical than using a properly normalized $g_2$, which demands only $20\langle\tau\rangle$, as shown in Figure~\ref{fig:MD_g2_comp}. Thus, we can expect that studies that utilized $\tilde{g}_2$ to capture the structural dynamics with a total experiment duration shorter than $40\langle\tau\rangle$ will underestimate $\tau$ and overestimate $\beta$.

For instance, in Zhang et al.~\cite{Zhang2018}, Chatterjee et al.~\cite{Chatterjee2021}, and Huang and Voyles~\cite{Huang2024}, the experiment duration spanned roughly 20 to 40 times the average relaxation time. Within this regime, while relaxation time and stretching exponent measurements may suffer from quantitative inaccuracy, the overarching insights remain qualitatively valid; that is, regions characterized by faster dynamics (such as free surfaces) consistently yield smaller estimated $\tau$ values compared to more sluggish domains. In contrast, the relative experimental durations utilized by Vaerst et al.~\cite{Vaerst2023} and Spangenberg et al.~\cite{Spangenberg2021} reached only 4 to 6 times the reported slow relaxation times. Such constrained observation windows introduce two critical complications. First, the true relaxation times are likely to be significantly greater than the published values, potentially compromising even qualitative trends. As demonstrated in Figure~\ref{fig:MD_g2_comp}(b), the $\tau$ values extracted via $\tilde{g}_2$ depend heavily on the relative duration of data collection within this window. Under a fixed experimental runtime, domains with shorter actual relaxation times correspond to a longer relative experiment duration, thereby suffering less severe underestimation. It is unclear whether this rapid, non-linear variation in underestimation severity can ultimately invert the perceived order of local dynamics. Second, the stretching exponent $\beta$ are drastically overestimated. Both Vaerst et al. and Spangenberg et al. documented $\beta$ values larger than 1 for multiple measurements. Physically, a compressed exponential regime ($\beta>1$) implies a fundamentally distinct state compared to standard stretching exponential state ($\beta\leq 1$)~\cite{Ruta2012,Fluerasu2007}. However, Figure~\ref{fig:MD_g2_comp}(b) shows that at this range of relative experiment durations, $\beta$ parameters can be artificially driven above 1 even when the system is undergoing a true $\beta < 1$ stretched exponential decay. Consequently, detecting compressed-like dynamics ($\beta>1$) within a $\tilde{g}_2$ framework may warrant more careful examination.

\begin{figure}[H]
\centering
\makebox[\textwidth][c]{\includegraphics{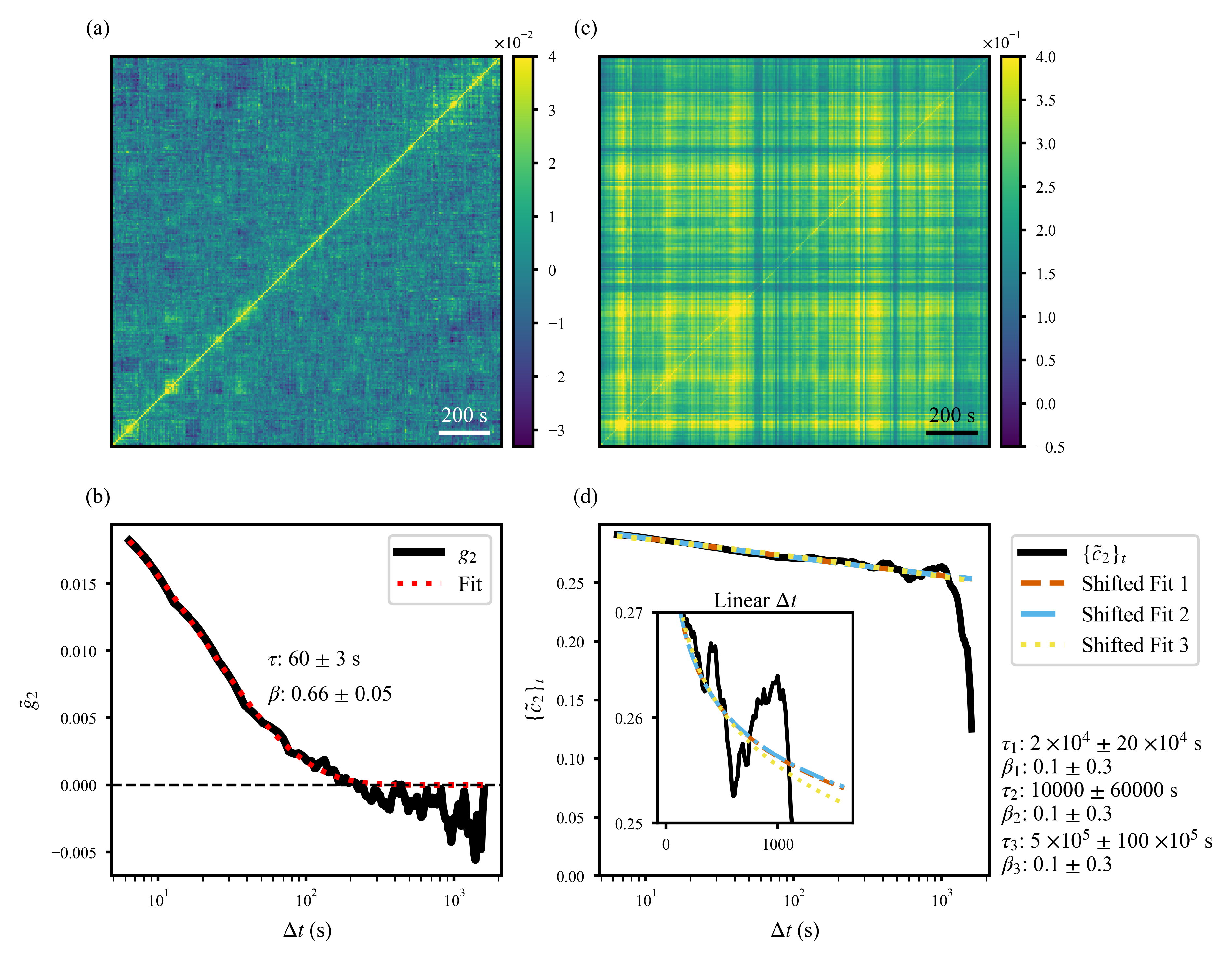}}
\caption{(a,b) Local TTCF and $\tilde{g}_2$ of crystal 1 marked in Figure~\ref{fig:Pt_mtau}, where the TTCF is calculated by approximating $\overline{I}=\{ I \}_{t_w}$; (c,d) local TTCF and $\{ \tilde{c}_2 \}_{t_w} $ from the same domain, using $\overline{I}=\langle I \rangle_{\mathbf{k}}$. The $\{ \tilde{c}_2 \}_{t_w} $ in (d) is fitted using Equation~\ref{eq:shifted_scaled_KWW} with three different initial guesses (0, 0.5, and 0.8) of $D$. The inset shows the original and fitted functions in linear $\Delta t$. The three initial guesses result in almost identical fitting curves within the experiment $\Delta t$ range, while the slowest and fastest estimated relaxation times vary by a factor of 50.  }
\label{fig:Pt_discussion}
\end{figure}

While $\tilde{g}_2$ might be serviceable when the experiment duration is relatively long compared to the global relaxation time, it can still fail catastrophically when local domains with extremely slow dynamics compared to the rest of the sample occur, be it crystals in SCLs or stable clusters in glasses. Because $\tilde{g}_2$ is constrained to drop to 0 even if the structure never changes, it does not detect such unchanging structures. Figure~\ref{fig:Pt_discussion} (a) and (b) show the TTCF from crystal 1 (shown in Figure~\ref{fig:Pt_crystals}) calculated using the intensity normalization $\delta I=(I-\{ I \}_{t_w})/\{ I \}_{t_w}$, which is the same normalization as $\tilde{g}_2$, and $\tilde{g}_2$. Even though the crystal never relaxed, $\tilde{g}_2$ still produced an estimated structural relaxation time of 60 seconds. Because of this issue, when Huang and Voyles first analyzed this dataset~\cite{Huang2024}, the crystallization events were not detected and erroneous dynamical information from the crystals was mixed into the liquid dynamics analysis. We suspect that some crystals were also present in Chatterjee et al.'s work on the same material using tilted-DF ECM, since regions that resemble crystals are observed in the tilted-DF image. Due to the large number of time series collected in an ECM experiment at different $\mathbf{r}$ and $\mathbf{k}$, these routines are typically automatically applied to the entire dataset without manual examination, potentially leading to extremely long lived domains being erroneously assigned short relaxation times without being noticed.

Some researchers have suggested that the $\mathbf{k}$-averaged intensity normalized two-time correlation function $\tilde{c}_2$ requires a shorter experiment duration compared to the one-time correlation function $\tilde{g}_2$~\cite{Nakazawa2023Structure,Nakazawa2025}. Nakazawa and Mitsuishi~\cite{Nakazawa2025} claimed that they were able to accurately estimate structural relaxation times longer than the entire experiment duration using $\tilde{c}_2$ calculated from NBED-ECM dataset in metallic glasses by applying a specific version of the shifted KWW fit, 
\begin{equation}
    \{ \tilde{c}_2 \}_{t_w} =C(1-D)\exp\left[-2\left(\frac{\Delta t}{\tau}\right)^\beta\right] + CD, \label{eq:shifted_scaled_KWW}
\end{equation}
where $C$ and $D$ are fitting parameters. Although they showed the efficacy of this shifted KWW fit on retrieving the structural dynamic parameters from a series of synthetic $\{ \tilde{c}_2 \}_{t_w} $ curves, we believe that applying this approach to actual NBED-ECM data with short relative experiment durations can be problematic. First, because systematic intensity variations across $k_r$ shift every $\tilde{c}_2$ upward, genuine structural stability cannot be disentangled from baseline correlation artifacts. Consequently, estimating $\tau$ depends heavily on the initial guess for the background shift parameter $D$. For example, Figures~\ref{fig:Pt_discussion}(c) and (d) present the $\tilde{c}_2$ TTCF from crystal 1 alongside the corresponding $\{ \tilde{c}_2 \}_{t_w}$. Fitting via Equation~\ref{eq:shifted_scaled_KWW} yields relaxation times of $(2\pm 20) \times 10^4$, $(1\pm 6) \times 10^4$, and $(50\pm1000) \times 10^4$ seconds for initial guesses of $D=0$, 0.5, and 0.8, respectively. The additional free parameter allows drastically different relaxation times to yield nearly identical fits within the experimental $\Delta t$ window. Moreover, because fitting uncertainties exceed the estimated relaxation times themselves, applying a shifted KWW fit to extremely slow domains in NBED-ECM data is fundamentally unreliable.

Second, even if the actual structural correlation can be properly separated from the constant background of $\{ \tilde{c}_2 \}_{t_w} $, there is no reason to assume that any single relaxation event should obey the chosen shifted KWW form. It is crucial not to over-extrapolate the autocorrelation function from a short experiment to estimate relaxation times for extremely slow or immobile domains, especially in NBED-ECM. In XPCS, the simultaneous sampling of multiple domains allows the relaxation process to be captured through ensemble averaging, even if individual domains do not complete a relaxation cycle during the observation time. In contrast, the nanometer-scale probe used in NBED-ECM often probes a single domain. If a domain does not relax during the observation time, estimating its relaxation time is highly speculative. In this sense, both the $1.8\times 10^7$ seconds relaxation time estimated from $g_2$ for the Pt\textsubscript{57.5}Cu\textsubscript{14.7}Ni\textsubscript{5.3}P\textsubscript{22.5} nanowire, and the 10000 seconds relaxation time estimated by the $\{ \tilde{c}_2 \}_{t_w}$ with background fitting are equally inaccurate. However, the extremely high relaxation time estimated from $g_2$ remains physically meaningful as it signifies that the observed domain is nearly immobile on the experimental time scale. 

In general, we propose that the properly normalized $g_2$ should always be used for analyzing NBED-ECM data, since it is capable of detecting unchanging structures within a system and does not require any background fitting. As shown in Figure~\ref{fig:MD_g2_comp} (b), $g_2$ only requires a relative experiment time of $20 \tau$ to produce consistent relaxation times. For samples with long relaxation times or unstable samples, where long experiments are impractical, relative experiment durations between 5 to 20 $\tau$ remain viable. Within this intermediate range, the relaxation time estimations only show a slight dependence on experiment duration, while the evaluation of the stretching exponent $\beta$ remains highly reliable and accurate throughout. Experiments shorter than $5\langle\tau\rangle$ are likely to be unreliable.

\subsection{Comparing XPCS and ECM TTCF }

\begin{figure}[H]
\centering
\makebox[\textwidth][c]{\includegraphics{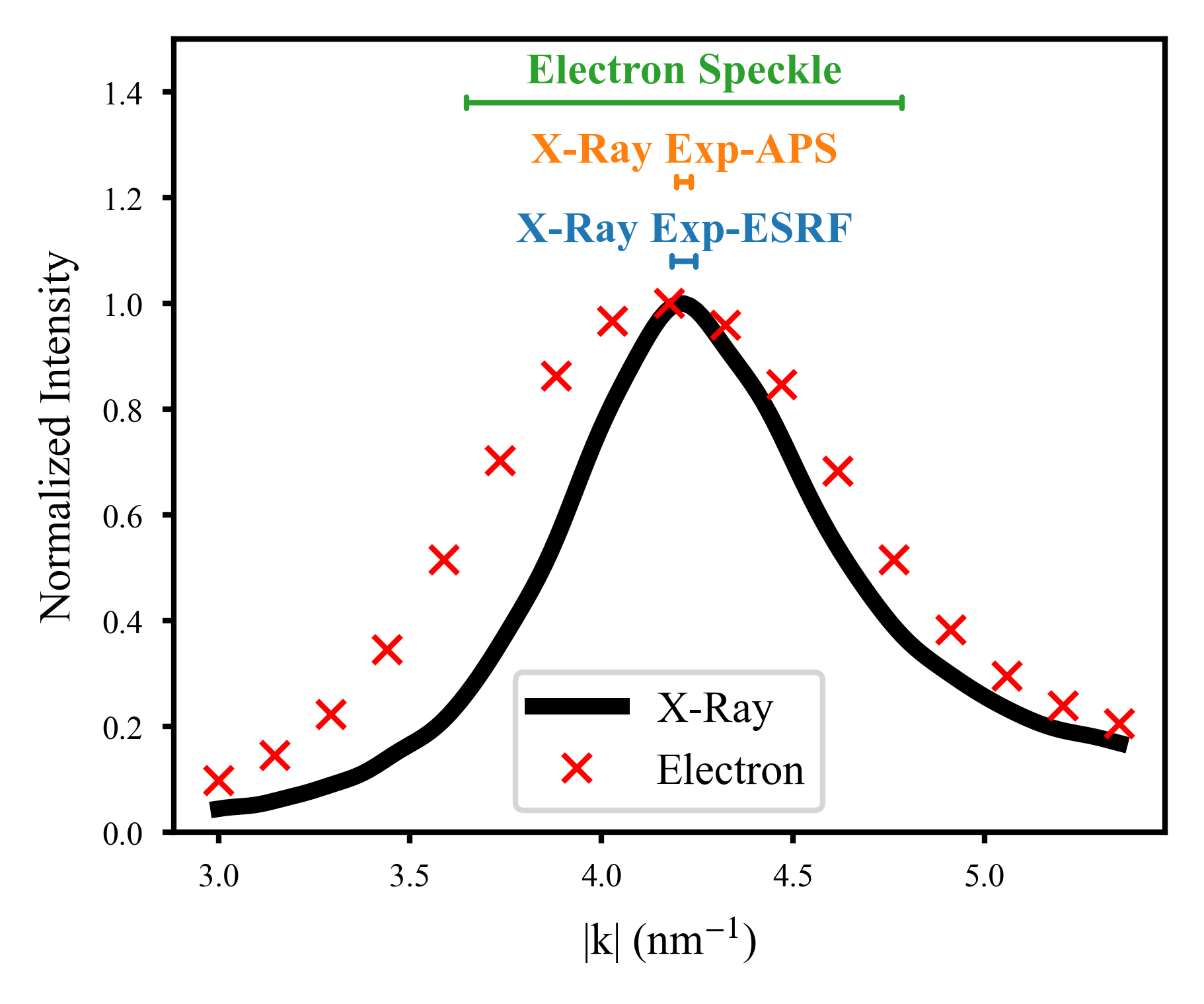}}
\caption[Simulated X-ray and electron scattering intensity profiles.]{Simulated X-ray and electron scattering normalized intensity profiles of the 760 K CuZr MD trajectory snapshot across the same $k_r$ range used in simulating the NBED-ECM data. The X-Ray Exp-APS and Exp-ESRF indicate the common metallic glasses XPCS experiment $k_r$ range used in the Advanced Photon Source (APS)~\cite{Das2020,Das2019,Riechers2024} and the European Synchrotron Radiation Facility (ESRF)~\cite{Ruta2012,Giordano2016,Evenson2015}, respectively. All X-ray ranges were shifted to center the CuZr X-ray intensity profile peak and transformed from units of $q=2\pi k$ (\AA$^{-1}$) to $k$ (nm$^{-1}$). The segment labeled Electron Speckle indicates the diameter of the diffraction disc of a 1.5 mrad convergent electron probe.}
\label{fig:X_ray}
\end{figure}

Because the constant shift in $\{ \tilde{c}_2 \}_{t_w} $ is caused by the variation of time-averaged intensity within the sampled $\mathbf{k}$ range, it is sensitive to the experimental $k_r$ range. If a method only samples a very small range of $k_r$, which is the case in many XPCS experiments~\cite{Malik1998,Livet2001,Sutton2003,Das2019,Das2020,Ruta2012,Giordano2016,Evenson2015,Riechers2024}, the shift in $\tilde{c}_2$ can be negligible. Figure~\ref{fig:X_ray} shows the simulated X-ray (7.32 keV) and electron scattering intensity profiles of the 760 K CuZr SCL MD trajectory. The X-ray speckle patterns were calculated by the magnitude of the direct sum of the structure factor, $F(\mathbf{k}) = \sum_{n} f^n_x(\mathbf{k}) \exp[2\pi i (\mathbf{k}\cdot\mathbf{r}_n)]$ and $I(\mathbf{k}) = |F(\mathbf{k})|^2$, where $\mathbf{r}_n$ is the position of the $n$-th atom, and $f^n_x$ is the X-ray atomic scattering factor obtained from Waasmaier and Kirfel\cite{Waasmaier1995}. To reduce artifacts from the shape factor of the cubic simulation box, only a central sphere with a diameter of 6.6 nm in the MD trajectory was included in the summation. These speckle patterns were simulated and averaged over every snapshot of the trajectory, and then summed over the azimuthal angle. The common ranges of $k_r$ used to calculate $\{ \tilde{c}_2 \}_{t_w} $ and the TTCF in metallic glass XPCS experiments~\cite{Das2020,Das2019,Riechers2024,Ruta2012,Giordano2016,Evenson2015} are marked in the figure. Compared to the width of the first diffraction ring ($\Delta k_r \approx$ 2.5 nm$^{-1}$), the experimental XPCS range is much smaller ($\Delta k_r \approx$ 0.1 nm$^{-1}$). At this narrow range, there is almost no intensity change across $k_r$. However, some XPCS experiments still report elevated backgrounds that required shifted KWW fitting~\cite{Das2020,Das2019,Riechers2024,Zhang2023}. It remains unclear whether this background is caused by actual intensity variations across $k_r$ or by other sources like detector defects and non-uniformity.

While some previous NBED-ECM studies also sample only a very thin slice of the diffraction ring~\cite{Nakazawa2023Structure,Nakazawa2025}, where $\{ \tilde{c}_2 \}_{t_w} $ appeared without a large background, we argue that $\tilde{c}_2$ is not ideal for NBED-ECM data for several reasons. First, because of the convergent beam, the electron speckle size in NBED patterns is much larger than in X-ray diffraction. A small $k_r$ range will therefore fail to probe the time evolution of any single speckle in its entirety. Second, $\tilde{c}_2$ is normalized using only the $\mathbf{k}$-averaged intensity. In XPCS, approximating the expected intensity as $\overline{I}(k_r,t_w)=\langle I(k_r,k_\phi,t_w) \rangle_{k_\phi}$ can be acceptable; a large illumination area (typically \SI{10}{\micro\meter} $\times$ \SI{10}{\micro\meter}) samples most possible structural motifs within a single pattern. While time averaging still provides the most accurate baseline~\cite{Sutton2003}, it is often omitted in XPCS to simplify data processing. In contrast, NBED patterns capture only a few motifs at a time, meaning an average over $k_\phi$ no longer reflects the true overall structure factor at each $k_r$. To achieve a proper estimation of $\overline{I}$, some time averaging is necessary. Therefore, unless the overall structure factor itself evolves due to global temperature shifts or crystallization, NBED-ECM data should always be normalized by the time- and azimuthal averaged intensity. This means evaluating the $g_2$ and $c_2$ using $\delta I(k_r, k_\phi, t_w) = (I(k_r, k_\phi,t_w)-\langle \{ I(k_r,k_\phi,t_w) \}_{t_w}  \rangle_{k_{\phi\pm\Delta\phi}})/\langle \{ I(k_r,k_\phi,t_w) \}_{t_w}  \rangle_{k_{\phi\pm\Delta\phi}}$. Furthermore, because it probes only a few structural motifs at once, NBED-ECM is sensitive to individual short-range and medium-range ordering clusters in SCLs and glasses. Variations in the structure, chemical composition, and orientation of these clusters cause their diffraction signals to shift to different $k_r$ and $k_\phi$ all within the first diffraction ring. Sampling only the very center of the ring can result in a significant loss of structural information. As an extreme example, crystal 2 in the Pt\textsubscript{57.5}Cu\textsubscript{14.7}Ni\textsubscript{5.3}P\textsubscript{22.5} nanowire exhibits a strong diffraction peak near the outer edge of the ring, as shown in Figure~\ref{fig:Pt_crystals} (e). If the TTCF and $g_2$ were calculated using only the ring center, this crystallization event might have gone undetected. Because the primary goal of ECM is to track the temporal behavior of these local configurations, the experiment should, as a rule of thumb, sample the same $k_r$ range used to study the static structure of disordered materials, including ring intensity variance~\cite{Nakazawa2023Structure} and the symmetry coefficient~\cite{Huang2022}.

The sampled $k_r$ ranges in tilted-DF ECM typically fall between those of XPCS and NBED-ECM. However, because all scattering intensities within the objective aperture integrate to form the real space images, the artifacts driven by intensity variations across $k_r$ do not apply to tilted-DF ECM. Nevertheless, the insights gained from NBED-ECM remain highly relevant. In tilted-DF ECM, $\tilde{c}_2$ is normalized by the $\mathbf{r}$-averaged intensity: $\tilde{c}_2(t_w, \Delta t)=\langle \delta\tilde{I}(t_w)\delta\tilde{I}(t_w+\Delta t) \rangle_{\mathbf{r}}$, where $ \delta\tilde{I} = (I-\langle I \rangle_{\mathbf{r}})/\langle I\rangle_{\mathbf{r}}$. If systematic variations (such as sample thickness variations or the presence of nanocrystals) exist within the tilted-DF image field of view, $\{ \tilde{c}_2 \}_{t_w} $ will display an elevated background at high delay times. This baseline elevation causes the same fitting issues when extracting $\tau$ and $\beta$ via the KWW model as seen in NBED-ECM. Consequently, extra precautions must be taken during tilted-DF ECM experiments to minimize these structural variations, or custom intensity normalization methods must be applied.

\section{Conclusion}

In this work, we addressed a critical methodological issue in NBED-ECM. We demonstrated that conventional XPCS normalization techniques—namely time averaging ($\tilde{g}_2$) and $\mathbf{k}$-averaging ($\tilde{c}_2$)—suffer from systematic errors when applied to the nanometer sampling volumes of NBED-ECM, leading to inaccurate relaxation times ($\tau$) and stretching exponents ($\beta$).

Our proposed time- and azimuthal-averaged normalization scheme resolves these challenges. Validation against MD-simulated CuZr SCL confirms the accuracy of our $g_2$ and $c_2$ frameworks across various observation window lengths. Furthermore, when applied to experimental 4D STEM datasets of a metallic glass nanowire, our methodology successfully isolated ultra-stable nanoscale crystal phases. While previous normalization techniques miscategorized these rigid domains as having short-lived fluid relaxations, our approach correctly preserves their immobile character on the experimental timescale. This ability to capture individual relaxation events highlights the unique capability of time-resolved 4D STEM-based ECM. With a proper experimental setup and the correct analysis, this technique allows the direct observation of mobile domains' size, distribution, lifetime, structure, and interaction with their surroundings, leading to a better understanding of structural relaxation and dynamic heterogeneity in SCLs. Ultimately, avoiding arbitrary background corrections by using these proper spatial and temporal references provides a reliable, artifact-free methodology for automated time-resolved 4D STEM data pipelines to accurately evaluate local spatiotemporal dynamics.

\appendix

\section{Conventional Definition of One-Time and Two-Time Correlation Functions}\label{append:acf_def}

Conventionally, the one-time correlation function and the KWW fit are expressed as:
\begin{equation*}
    G_2(\mathbf{k},\Delta t) = \frac{\{I(\mathbf{k},t_w)I(\mathbf{k},t_w+\Delta t)\}_{t_w}}{\overline{I}^2} = 1 + A\exp\left[-2\left(\frac{\Delta t}{\tau}\right)^\beta\right]
\end{equation*}
Thus, the decaying part of $G_2$ is:
\begin{align*}
    g_2(\mathbf{k},\Delta t) &= G_2(\mathbf{k}, \Delta t) - 1 = \frac{\{I(\mathbf{k},t_w)I(\mathbf{k},t_w+\Delta t)\}_{t_w}-\overline{I}^2}{\overline{I}^2}, \\
    &= \frac{\{[I(\mathbf{k},t_w)-\overline{I}][I(\mathbf{k},t_w+\Delta t)-\overline{I}]\}_{t_w}}{\overline{I}^2}, \\ 
    &= \left\{\left[\frac{I(\mathbf{k},t_w)-\overline{I}}{\overline{I}}\right]\left[\frac{I(\mathbf{k},t_w+\Delta t)-\overline{I}}{\overline{I}}\right]\right\}_{t_w},
\end{align*}
which has the same form as Equation~\ref{eq:g2}.

The two-time correlation function is conventionally expressed as:
\begin{equation*}
    C_2(t_w,\Delta t) = \frac{\langle I(\mathbf{k},t_w)I(\mathbf{k},t_w+\Delta t)\rangle_{\mathbf{k}}}{\overline{I}(t_w)\overline{I}(t_w+\Delta t)} = 1 + A\exp\left[-2\left(\frac{\Delta t}{\tau_m}\right)^{\beta_m}\right]
\end{equation*}
Thus, the decaying part of $C_2$ is:
\begin{equation*}
    c_2(t_w,\Delta t) = C_2(t_w, \Delta t) - 1 = \frac{\langle I(\mathbf{k},t_w)I(\mathbf{k},t_w+\Delta t)\rangle_{\mathbf{k}}-\overline{I}(t_w)\overline{I}(t_w+\Delta t)}{\overline{I}(t_w)\overline{I}(t_w+\Delta t)}.
\end{equation*}
Assuming $\langle I(\mathbf{k}, t_w)\rangle_\mathbf{k}\approx \langle I(\mathbf{k}, t_w+\Delta t)\rangle_\mathbf{k}$ for a stable system, then:
\begin{align*}
    c_2(t_w,\Delta t)&= \frac{\langle[I(\mathbf{k},t_w)-\overline{I}][I(\mathbf{k},t_w+\Delta t)-\overline{I}]\rangle_{\mathbf{k}}}{\overline{I}(t_w)\overline{I}(t_w+\Delta t)}, \\ 
    &= \left\langle\left[\frac{I(\mathbf{k},t_w)-\overline{I}(t_w)}{\overline{I}(t_w)}\right]\left[\frac{I(\mathbf{k},t_w+\Delta t)-\overline{I}(t_w+\Delta t)}{\overline{I}(t_w+\Delta t)}\right]\right\rangle_{\mathbf{k}}, \\
    &= \langle \delta I(\mathbf{k},t_w) \delta I(\mathbf{k},t_w+\Delta t)\rangle_\mathbf{k}
\end{align*}
which has the same form as Equation~\ref{eq:c2}, where $\delta I(\mathbf{k},t_w) = (I(\mathbf{k},t_w)-\overline{I})/\overline{I}$.

\section{Origin of the High \texorpdfstring{$\{\tilde{c}_2\}_{t_w}$}{} in NBED-ECM} \label{append:c2}

Assuming that the system is in equilibrium, then we can approximate $\langle I(t_w) \rangle_{\mathbf{k}}\approx \langle I(t_w+\Delta t) \rangle_{\mathbf{k}}\approx \{ \langle I \rangle_{\mathbf{k}}  \}_{t_w} \approx C$, where C is a constant. If we define $\delta \tilde{I} = (I-\{ I \}_{t_w})/\{ I \}_{t_w}$ and $I=\delta \tilde{I}\{ I \}_{t_w} + \{ I \}_{t_w} $, then $\tilde{c}_2 $ becomes:
\begin{align*}
    \tilde{c}_2 &= \left\langle \frac{I(t_w)-\langle I(t_w) \rangle_{\mathbf{k}} }{\langle I(t_w) \rangle_{\mathbf{k}} }\frac{I(t_w+\Delta t)-\langle I(t_w+\Delta t) \rangle_{\mathbf{k}} }{\langle I(t_w+\Delta t) \rangle_{\mathbf{k}} } \right\rangle_{\mathbf{k}}, \\
    &= \left\langle \left(\frac{\delta \tilde{I}(t_w)\{ I \}_{t_w}+\{ I \}_{t_w}  -C}{C}\right)\left(\frac{\delta \tilde{I}(t_w+\Delta t)\{ I \}_{t_w} +\{ I \}_{t_w} - C}{C}\right) \right\rangle_{\mathbf{k}}, \\
    &= \frac{1}{C^2}\left\langle (\delta \tilde{I}(t_w)\{ I \}_{t_w}+\{ I \}_{t_w}  -C)(\delta \tilde{I}(t_w+\Delta t)\{ I \}_{t_w}+\{ I \}_{t_w}  -C) \right\rangle_{\mathbf{k}}, 
\end{align*}
where the $\mathbf{k}$ dependence in $\delta\tilde{I}(\mathbf{k},t_w)$ and $\{ I(\mathbf{k}) \}_{t_w} $ are omitted for simplicity. Since $\{ \delta \tilde{I} \}_{t_w} =0$, then:
\begin{align*}
    \{ \tilde{c}_2 \}_{t_w} &= \frac{1}{C^2}\left\{ \left\langle (\delta \tilde{I}(t_w)\{ I \}_{t_w}+\{ I \}_{t_w}  -C)(\delta \tilde{I}(t_w+\Delta t)\{ I \}_{t_w}+\{ I \}_{t_w}  -C) \right\rangle_{\mathbf{k}} \right\}_{t_w}, \\
    &= \frac{1}{C^2}\left\langle \left\{ (\delta \tilde{I}(t_w)\{ I \}_{t_w}+\{ I \}_{t_w}  -C)(\delta \tilde{I}(t_w+\Delta t)\{ I \}_{t_w}+\{ I \}_{t_w}  -C) \right\}_{t_w}  \right\rangle_{\mathbf{k}}, \\
    &= \frac{1}{C^2}\left\langle \left\{ \delta\tilde{I}(t_w)\delta\tilde{I}(t_w+\Delta t)\{ I \}_{t_w}^2 + \{ I \}_{t_w}^2 - 2C \{ I \}_{t_w} + C^2   \right\}_{t_w}  \right\rangle_{\mathbf{k}}, \\
    &= \left\langle \{ \delta\tilde{I}(t_w)\delta\tilde{I}(t_w+\Delta t) \}_{t_w} \frac{\{ I \}_{t_w}^2 }{C^2} \right\rangle_{\mathbf{k}} + \left\langle \frac{(\{ I \}_{t_w}-C )^2}{C^2} \right\rangle_{\mathbf{k}}. 
\end{align*}

If we include the $\mathbf{k}$ dependence back in the equation and examine the first term, we can find:
\begin{align*}
    \left\langle \{ \delta\tilde{I}(\mathbf{k},t_w)\delta\tilde{I}(\mathbf{k},t_w+\Delta t) \}_{t_w} \frac{\{ I(\mathbf{k}) \}_{t_w}^2 }{C^2} \right\rangle_{\mathbf{k}} &= \left\langle \tilde{g}_2(\mathbf{k}) \frac{\{ I(\mathbf{k}) \}_{t_w}^2 }{\langle \{ I \}_{t_w}  \rangle_{\mathbf{k}}^2 }\right\rangle_{\mathbf{k}}, 
\end{align*}
which is a weighted $\mathbf{k}$-averaged $\tilde{g}_2$ function. And the second term becomes:
\begin{align*}
    \left\langle \frac{(\{ I \}_{t_w}-C )^2}{C^2} \right\rangle_{\mathbf{k}} &= \frac{\langle \{ I(\mathbf{k}) \}_{t_w} - \langle \{ I \}_{t_w}  \rangle_{\mathbf{k}}  \rangle_{\mathbf{k}} }{\langle \{ I \}_{t_w}  \rangle_{\mathbf{k}}^2 }, \\
    &= \frac{\mathrm{Var}_{\mathbf{k}}[\{ I \}_{t_w} ]}{\langle \{ I \}_{t_w}  \rangle_{\mathbf{k}}^2},
\end{align*}
where $\mathrm{Var}_\mathbf{k}[\{ I \}_{t_w} ]$ is the intensity variance of time-averaged diffraction pattern. 

Therefore, the $\{ \tilde{c}_2 \}_{t_w} $ function can be approximated as the weighted $\mathbf{k}$-averaged $\tilde{g}_2$ function plus the normalized variance of every pixel in the time-averaged diffraction pattern. If the diffraction ring pattern include a large $k_r$ range, than the normalized ring variance of $\{ I \}_{t_w} $ will be large due the the variation in structure factor across the ring.

The weighted $\mathbf{k}$-averaged $\tilde{g}_2$ function might have a slightly different shape than regular $\langle \tilde{g}_2 \rangle_{\mathbf{k}} $ function, but they share the same properties. Thus we can analyze the limiting behaviors of $\{ \tilde{c}_2 \}_{t_w} $ at long and short experiment duration. When the experiment duration is extremely long compared to the structural relaxation time, the $\tilde{g}_2$ function will approach 0 at high delay time, thus the autocorrelation function will reach the normalized time-averaged ring variance, which is the origin of the artificially high autocorrelation in $\{ \tilde{c}_2 \}_{t_w}$. On the other extreme, if the experiment duration is very short, the $\tilde{g}_2$ function will become very negative, canceling out the normalized time-averaged ring variance, lowering the estimated relaxation time. This is the reason why at short time-series length, we can observe the estimated $\tau$ from $\{ \tilde{c}_2 \}_{t_w} $ function to decrease in Figure~\ref{fig:MD_g2_comp} (b).

\begin{acknowledgements}
    The authors thank Shuoyuan Huang for making metallic glass nanowire ECM data available. During the preparation of this work, large language model (Google Gemini 3.5) was used strictly for language polishing, grammar correction, and improving readability. The output of the model was reviewed and edited by the authors as needed.
\end{acknowledgements}

\begin{funding}
    This work is supported by the Wisconsin Materials Research Science and Engineering Center (MRSEC) DMR-2309000. The use of Wisconsin Centers for Nanoscale Technology facilities and instrumentation is also supported by MRSEC DMR-2309000. 
\end{funding}

\ConflictsOfInterest{
The authors declares no conflicts of interest.
}

\DataAvailability{
The data and scripts supporting this study are deposited in the Figshare repository with DOI: 10.6084/m9.figshare.33198822 and are available from the corresponding authors upon reasonable request.}

\bibliography{reference}

\end{document}